**Spin crossover self-assembled monolayer of a Co(II) terpyridine derivative functionalized with carboxylic acid groups**

*Víctor García-López, Niccolò Giaconi, Lorenzo Poggini, Amelie Juhin, Brunetto Cortigiani, Javier Herrero-Martín, Matteo Mannini,* Miguel Clemente-León* and Eugenio Coronado*

Víctor García-López, Miguel Clemente-León* and Eugenio Coronado
Instituto de Ciencia MolecularUniversidad de ValenciaCatedrático José Beltrán 2, Paterna 46980, Spain
E-mail: miguel.clemente@uv.es
Víctor García-López, Niccolò Giaconi, Brunetto Cortigiani, Matteo Mannini*
Department of Chemistry "Ugo Schiff" and INSTM Research Unit, University of Florence, Via della Lastruccia 3–13, 50019 Sesto Fiorentino (Italy)
E-mail: matteo.mannini@unifi.it
Lorenzo Poggini
Istituto di Chimica dei Composti Organometallici (ICCOM), CNR Via Madonna del Piano, 10, 50019 Sesto Fiorentino (Italy)
Amélie Juhin
Sorbonne Université, UMR CNRS 7590, MNHN, Institut de Minéralogie, de Physique des Matériaux et de Cosmochimie, IMPMC, 75005, Paris, France
Javier Herrero-Martín
ALBA Synchrotron, Carrer de la Llum 2-26, 08290 Cerdanyola del Vallès, Barcelona, Spain

## Abstract

The deposition of a monolayer by direct self-assembly from solution of an active spin-crossover (SCO) complex via chemisorption has been achieved starting from the perchlorate salt of Co(II) bis(4´-(4-carboxyphenyl)-2,2':6',2''-terpyridine). MALDI-TOF MS, Raman spectroscopy and X-ray photoelectron spectroscopy (XPS) confirm the presence of an intact monolayer of SCO molecules grafted through carboxylate groups to the Ag surface. Three different spectroscopic techniques (Raman, XPS and X-ray Absorption spectroscopy, XAS) evidence a thermal spin transition of such monolayer of molecules. To our knowledge, this is the first example of an active SCO assembly in direct contact with a surface obtained by wet chemistry.

## 1. Introduction

Being able to control bistable nanostructures at the molecular level assures a breakthrough in memory and data storage devices.[1] In this regard, spin-crossover (SCO) complexes showing magnetic bistability, also at room temperature,[2–4] have attracted lots of attention since both, low spin (LS) and high-spin (HS) states are reachable through external stimuli such us temperature, pressure or light.[5–7] Deposition of the complexes on solid surfaces is a mandatory step for applications such as electrical switches[8–10] or mechanical actuators[11]. Over the past ten years, the preferred method for depositing this type of systems has been sublimation in ultrahigh vacuum of neutral complexes.[12–16] Actually, it has been proven that nanostructured assemblies achieved in ultra-high vacuum of SCO systems can maintain switchability even down to the (sub)monolayer regime.[17–19] However, this has restricted this deposition method to a few families of vacuum evaporable molecules.[20] An interesting alternative to expand the number of deposited complexes is self-assembly of properly functionalized molecules from solution. However, the preparation of self-assembled monolayers (SAMs) of SCO molecules have remained almost unexplored, excluding some earlier reports regarding chemisorption of such molecules in a junction-like gold nanoparticle array.[21] This soft and cost-effective deposition method is a general simple grafting method that only requires molecules (neutral or not) functionalized with suitable anchoring groups capable of specifically interact with the surface.[22] This approach has been already employed with SCO complexes to be part of an interfacial device,[23] or with valence-tautomerism switching molecules.[24] While previous attempts to form SAMs of Fe(II) SCO systems were unsuccessful,[25–27] the Co(II)-based SCO complex chosen in this work have been preferred due to its robustness against oxidation compared to the Fe(II)-based.[25] Spin transition in six-coordinated Co(II) compounds ($d^7$) involves a spin conversion from a S = 1/2 ($t_{2g}^6 e_g^1$) LS state to a S = 3/2 ($t_{2g}^5 e_g^2$) HS state. Many bis-chelated Co(II) SCO complexes based on tridentate ligands of 2,2’:6’,2’’-terpyridine (terpy) derivatives have been reported.[28] Among them, those containing substituents at the C(4′) position are of special interest.[29] Indeed, it has been shown that different substituents allow diverse magnetic properties ranging from abrupt spin transitions to unique SCO behaviors such as reverse spin transitions.[30] Furthermore, the use of specific substituents has enabled to combine SCO with other properties such as porosity,[31] liquid crystal behavior,[32–35] ferroelectricity[36] or gas sensing.[37] In this work, we have taken profit of its ability to carry additional features while maintaining SCO by using a Co(II) complex based on terpy derivative bearing a carboxylic acid group (4´-(4-carboxyphenyl)-2,2':6',2''-terpyridine, **HL**),[38] see **Figure 1**. Hereby, we intend to take advantage of the anchoring capabilities of the carboxylic acid group on noble-metal surfaces to form SAMs from solution of a Co(II) SCO complex.

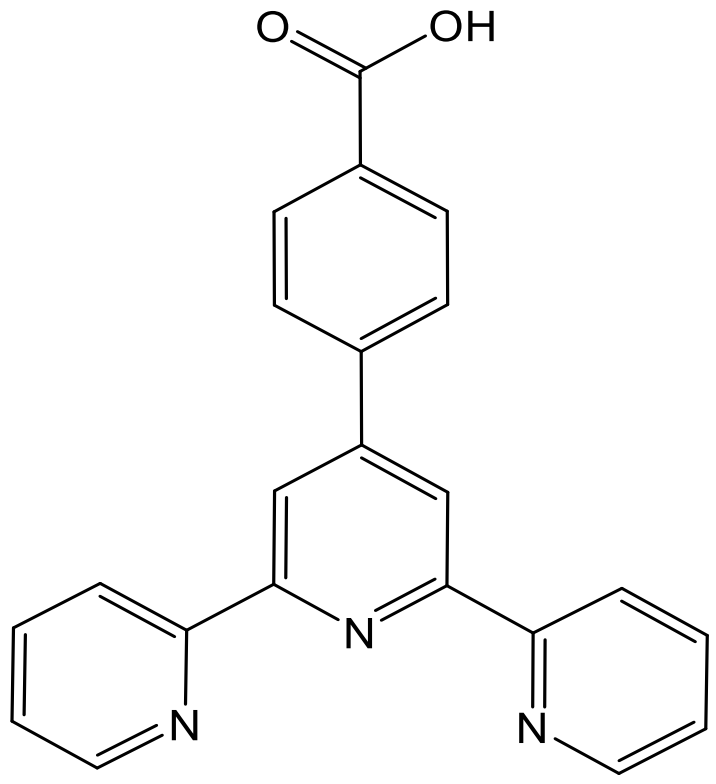

**HL**

**Figure 1.** Molecular structure of the ligand 4´-(4-carboxyphenyl)-2,2':6',2''-terpyridine (**HL**).

The structure of the Co(II) neutral complex with the deprotonated ligand **L**, $[Co(L)_2]\cdot 5H_2O$, was reported in the literature by Yang *et. al.*,[39] but the magnetic properties were not studied. In this work, we have prepared the complex of the protonated ligand in order to optimize its processability for deposition on surfaces. Thus, we prepared the $[Co(LH)_2]\cdot(ClO_4)_2\cdot 4DMA$ (**1**) compound, which has been structurally and magnetically characterized in the solid state, and we studied its behavior upon surface deposition. The morphology, stability, and composition of deposited monolayers of **1** on top of Ag (**1·SAM**) were studied by atomic force microscopy (AFM), matrix assisted laser desorption ionization - time-of-flight mass spectrometry (MALDI-TOF), X-ray photoelectron spectroscopy (XPS), and Raman spectroscopy. SCO properties were further investigated by means of X-ray absorption spectroscopy (XAS). XPS, Raman and XAS confirm that the spin state of the Co(II) complexes on top of Ag displayed gradual and reversible changes with temperature. Therefore, it represents, as far as we are aware, the first example of an active SCO SAM from solution in direct contact with the surface.

## 2. Results and discussion

### 2.1 Solid state

#### *2.1.1. Synthesis*

**1** was obtained by slow diffusion of diethyl ether into solutions of $Co(ClO_4)_2\cdot xH_2O$ and **HL** in a 1:2 molar ratio in dimethylacetamide (DMA). After a few weeks, big prismatic dark-orange crystals suitable for single crystal diffraction were obtained. Purity and stability of the complex was checked with elemental analysis and powder X-ray diffraction (PXRD, **Figure S.1** and associated text in the Supporting Information).

*2.1.2. Structure*

The crystal structure of **1** was solved by single-crystal X-ray diffraction at 120 K, 300 K and 340 K in the non-centrosymmetric *P*-1 space group. The asymmetric units is composed of one $[Co(LH)_2]^{2+}$ cation, two $ClO_4^-$ anions and four DMA solvent molecules (**Figure 2a** and SI for more information). The metal center is coordinated to the six N of the two terpy moieties giving rise to a distorted octahedral geometry. Co-N bond lengths display a subtle increase in bond lengths with temperature (see **Table S.1** in the Supporting Information). This indicates that the number of molecules in the HS state is increasing with temperature, as it has been already seen in other compounds,[5] and agrees with magnetic properties (see below). Neighboring $[Co(LH)_2]^{2+}$ cations in the structure are weakly interacting through π···π and CH···π interactions (see **Figure S.2** in the Supporting Information), while hydrogen bonds with the DMA solvent molecules avoid the formation of 1D chains of hydrogen-bonded $[Co(LH)_2]^{2+}$ cations (see **Figure 2a**), which have been seen in other bis-tridentate 4-substituted pyridine derivatives.[40,41]

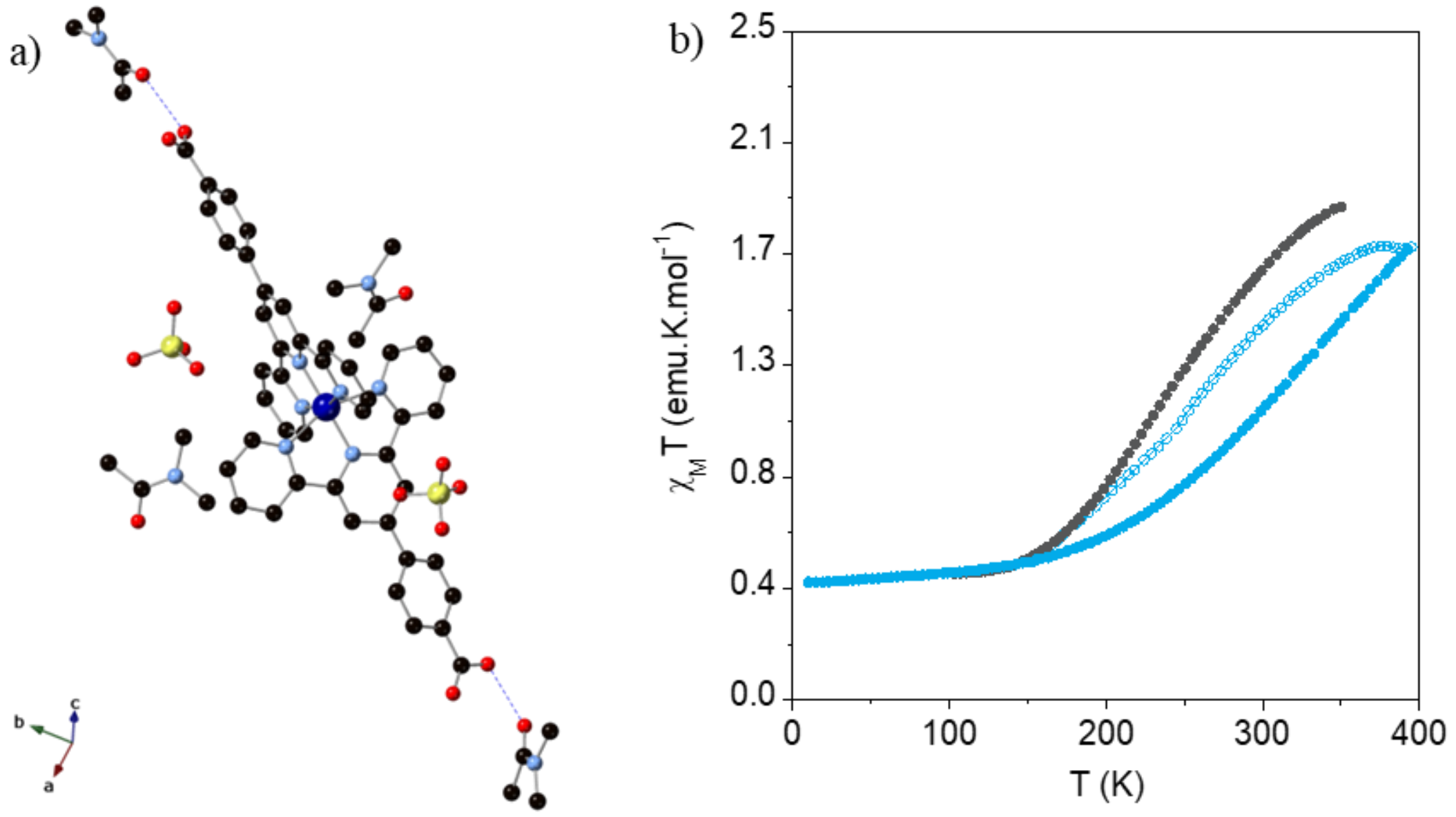

**Figure 2. a)** Structure of the complex **1** at 120 K. C (black), N (blue), O (red), Co (dark blue), hydrogen atoms are omitted for clarity, hydrogen bonds are represented with dashed blue lines. **b)** $\chi_M T$ vs temperature for **1** in contact with mother liquor (grey circles), filtered (blue empty circles) and after desolvation (filled blue circles).

*2.1.3. Magnetic properties*

The temperature dependence of $\chi_M T$ for **1** in contact with mother liquors presents a gradual increase from typical LS values below 150 K (0.48 $cm^3 \cdot K \cdot mol^{-1}$) to 1.85 $cm^3 \cdot K \cdot mol^{-1}$ at 350 K suggesting an incomplete and non-cooperative SCO transition (**Figure 2b**). Magnetic measurements on the filtered sample before and after heating to 400 K shows a similar behavior with a more incomplete SCO transition reaching a maximum value of 1.72 $cm^3 \cdot K \cdot mol^{-1}$ at 400K. This value corresponds to around 60% of molecules in the HS state, taking *ca.* 2.5 $cm^3 \cdot K \cdot mol^{-1}$ as the value for pure Co(II) in HS state.[30] In contrast, the Co(II) neutral complex with the deprotonated ligand **L**, $[Co(L)_2] \cdot 5H_2O$, shows an abrupt thermal spin transition upon dehydration (see **Figure S.3** and associated text in the Supporting Information). This suggests that the absence of strong intermolecular interactions between complexes in **1**, probably due to the presence of $ClO_4^-$ counteranions and DMA solvent molecules in the structure, could induce the gradual spin transition of **1**. On the other hand, the absence of these molecules in the dehydrated $[Co(L)_2]$ compound would involve closer contacts between complexes leading to a cooperative and abrupt SCO. A similar effect occurs in the Co(II) SCO complex based on the terpyridine derivative without the 1,4-phenylene spacer between the tpy and the carboxylic acid, the more compact structure obtained upon dehydration leads to a more abrupt SCO.[37]

## 2.2. Deposition

*2.2.1 Monolayer preparation*

**1·SAM**s were prepared by immersing evaporated silver substrates, after hydrogen flame clean-annealing (see Supporting Information for further details), for 6 h in a 1 mM anhydrous DMA solution of **1**. After that, they were washed and rinsed throroughly with clean DMA to remove any physiosorbed material and dried under $N_2$ stream. Silver was chosen due to the possibility to form SAMs through the carboxylate moiety[42] and because it has been reported that large domains of well-ordered 2D molecular arrays can be promoted through this kind of interaction at room temperature within a short time.[43]

*2.2.2. Atomic Force Microscopy (AFM)*

Topographic images of **1·SAM** confirmed the absence of aggregates or multilayers (see **Figure S.4** in the Supporting Information). This result was expected considering that the carboxylate group bonds selectively the first layer of molecules to the silver surface and additional physisorbed layers can be removed with the washing steps.

*2.2.3. Matrix Assisted Laser Desorption Ionization – Time-Of-Flight Mass Spectrometry (MALDI-TOF MS)*

Integrity of the molecules in **1·SAM** was checked with MALDI-TOF MS. A comparative analysis between bulk and SAM is shown in **Figure 3**. Peaks of *m/z* corresponding to $[Co(LH)_2]^+$ are present in bulk and monolayer spectra (764.16 *m/z*). Proof of the formation of a bond between only one carboxylate group of the molecules and the surface is supported by the presence of signals originated from a silver atom that is extracted together with molecular fragments such as $[L+H+Ag]^+$, [Co(LH+H+Na)(L+Ag)] and $[Co(LH\text{-}COOH+H)(L+Ag)]^+$; these signals are not present in the bulk sample (see **Figure 3**).

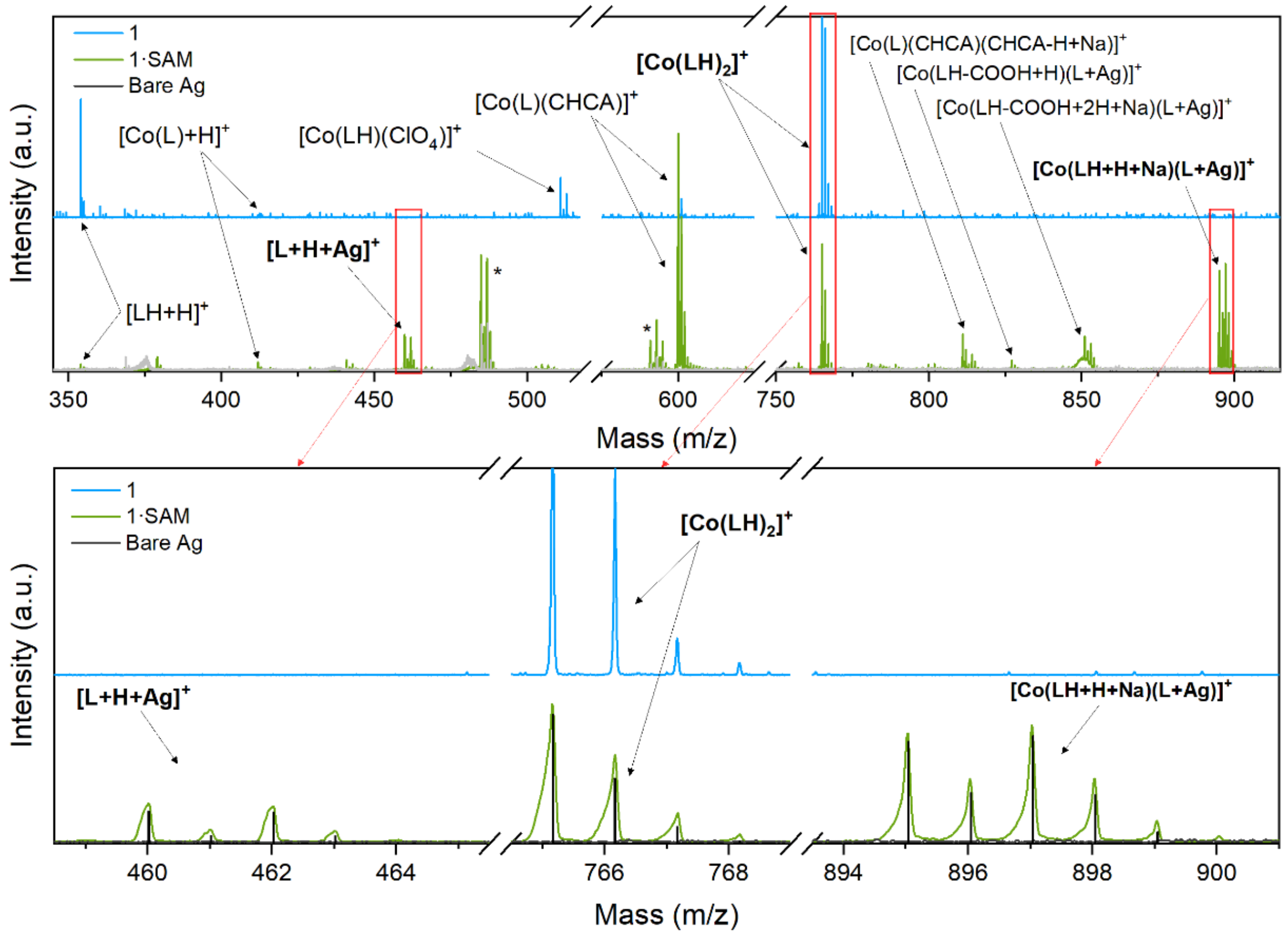

**Figure 3.** MALDI-TOF spectra of **1** (blue), **1·SAM** (green) and bare Ag (grey). Magnification of the $[L+H+Ag]^+$, $[Co(LH_2)]^+$ and $[Co(LH+H+Na)(L+Ag)]^+$ (enclosed in red rectangles). The expected isotopic distribution pattern for each fragment is reported as black lines. Peaks coming from the matrix used for calibration are marked with an asterisk and are also observed in bare Ag. CHCA is abbreviation for α-cyano-4-hydroxycinnamic acid and is present in the matrix used for MALDI-TOF MS measurements (see Experimental Section).

This result has also been observed previously in other systems assembled on Au,[24,44,45] where a molecule-substrate bonding is expected and suggests that deprotonation of the carboxylic acid group takes place upon contact with the surface. Additionally, the lack of molecular fragments in the monolayer sample with the counteranion, such us $[Co(LH)(ClO_4)]^+$, that are present in the bulk sample, also confirms the absence of aggregates or any physiosorbed multilayers in agreement with AFM measurements (for further information go to **Table S.3** in the Supporting Information).

### *2.2.4. Raman and SERS spectroscopy*

It was possible to study the deposits with Raman spectroscopy on top of a mechanically roughened silver substrate. The nature of the imposed silver defects and their small separation leads to a plasmon absorbance, which is resonant with the 532 nm light used to excite the samples.[46] Therefore, the Raman spectrum of the deposits is expected to be magnified in the "hot-spots" by the SERS effect.[47] The structural integrity of the molecules in **1·SAM** is evident from the fact that SERS spectra of the monolayer and conventional Raman spectra of the bulk are found to be very similar (**Figure 4**).

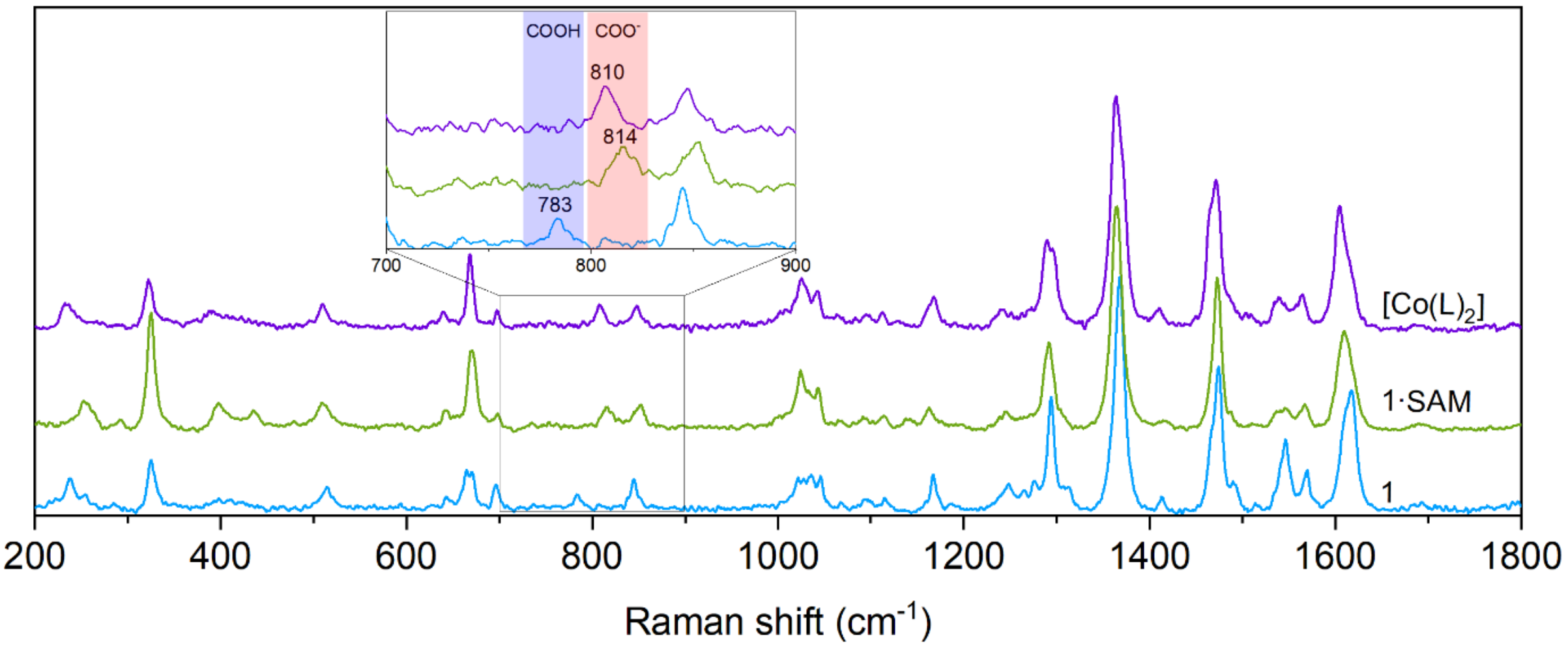

**Figure 4.** Raman fingerprints of **1** (blue), **1·SAM** (green) and **$[Co(L)_2]\cdot 5H_2O$** (purple) in the 200-1800 $cm^{-1}$ at 100K. Inset of the 700-900 $cm^{-1}$ region.

The most important difference between both samples is the replacement of the band at 783 $cm^{-1}$ in the bulk for another one at 814 $cm^{-1}$ in the monolayer. This band is assigned to the deformation vibration of carboxylate groups and is sensitive to the protonation/deprotonation state.[48] Such band is also present in the spectra of the complex $[Co(L)_2]\cdot 5H_2O$[39] (**Figure 4**) and in the spectra of a monolayer of the ligand (**L·SAM**, see **Figure S.7** in the Supporting Information), both measured as references. Deprotonation of the carboxylic acid group in contact with the surface was expected, as it has been already observed in literature experiments performed with similar ligands on Ag.[49] However, the protonation-deprotonation state of the carboxylic acid group non-interacting with the surface is not straightforward. Thus, the presence of only one band is indicative that both carboxylic acid groups of the deposited complexes are deprotonated,[50] which is consistent with the absence of $ClO_4^-$ counteranions observed in XPS experiments performed on **1·SAM** (see below). Additionally temperature dependent Raman spectroscopy experiments have been used to monitor the spin state variation of the Co(II). In the bulk sample, changes in intensity ratios and Raman shifts are observed at the same temperature range of the spin transition, especially in the coordinative *terpy* modes (1200-1700 $cm^{-1}$, **Figure S.6a**). Nevertheless, the most notably feature is the increase of intensity of the peak located at 1020 $cm^{-1}$ observed when increasing temperature. This peak is assigned to the pyridine ring breathing mode,[51] which is strongly coupled to the Co-N stretch and, in consequence, to the spin state. Similar Raman spectral changes of this peak accompanying spin transition are observed in a related Co(II) complex reported in the literature.[52] To get a more quantitative insight, the normalized intensity of such feature (integrated against the weak temperature-dependent mode of the 4-substituted position of the pyridine ring at 1365 $cm^{-1}$)[51] is plotted against temperature (**Figure S.6b**). It displays a curve in excellent agreement with the bulk data extracted from conventional magnetometry measurements. Therefore, these Raman spectral changes can be attributed to a spin transition and used to monitor the spin state. SERS spectra of **1·SAM** display similar changes with temperature (**Figure 5a**, for more temperatures go to **Figure S.7** in the Supporting Information). These changes are not observed in **L·SAM** (**Figure S.8**). Therefore, we can conclude that the spin state of **1·SAM** changes with temperature. A comparison of the spin transition of the bulk with that of the monolayers of **1** and **L,** using the same spin state-dependent marker, suggests that **1** and **1·SAM** undergo spin transition in the same temperature range (150-300 K), while **L·SAM** does not display any changes (**Figure S.8**). The fraction of HS molecules seems to be higher in **1·SAM**. These results are confirmed by XPS and XAS measurements, which enables a more accurate estimation of the LS/HS fraction (see below). To prove stability and reproducibility of the spin transition, after the first set of measurements, the sample was stored in air for one week and two more cooling and warming cycles were performed (see **Figure 5c** and **d**). They confirm that the spin transition in the monolayer is reversible in the temperature range 100 – 360 K. Further increase in temperature leads to irreversible changes in the spectra and blocking of the spin state (see **Figure S.9**). **L·SAM** exhibits the same irreversible changes (see **Figure S.10** and associated text) suggesting that the irreversibility of the spin transition at temperatures above 360 K is related to structural changes of the SAMs.[53,54] Indeed, the irreversibility of the spin transition when heating the SAMs above 360 K was confirmed by XAS (see below).

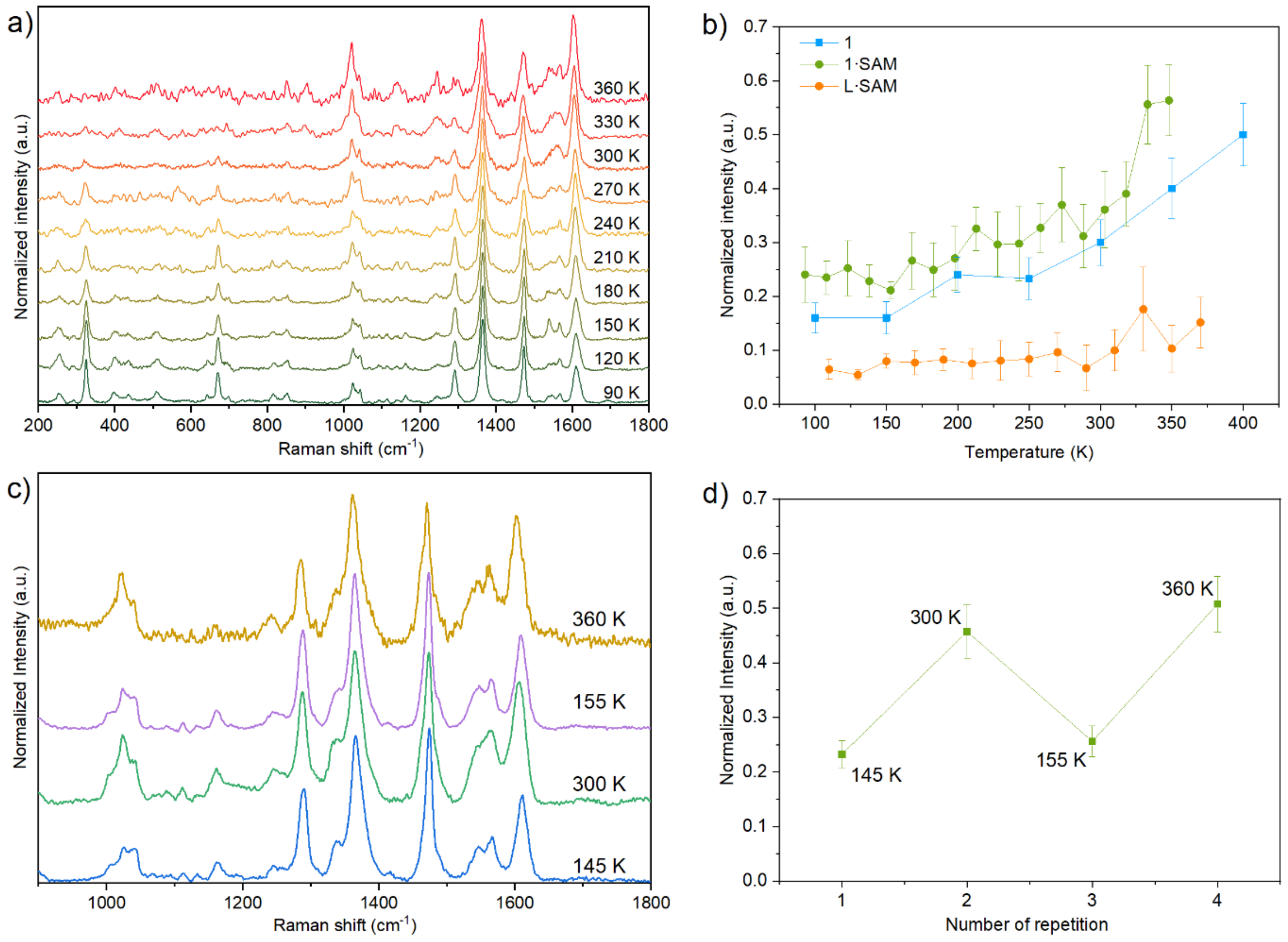

**Figure 5**. **a)** Temperature dependent Raman spectra of **1·SAM** in the 200-1800 cm$^{-1}$ range. **b)** Normalized intensity of the 1020 cm$^{-1}$ features for **1** (blue squares), **1·SAM** (green circles) and **L·SAM** (orange circles) for each temperature. **c)** Temperature dependent Raman spectra of the same sample after being stored in air for one week in the 900-1800 cm$^{-1}$ range. **d)** Values of the normalized intensity of the 1020 cm$^{-1}$ feature at each temperature for the subsequent cooling and warming cycles. The lines are a guide to the eye. Error bars have been calculated from the median absolute deviation.

### *2.2.5. X-ray Photoelectron spectroscopy*

C*1s*, N*1s*, Cl*2p* and Co*2p* XPS regions of **1** and **1·SAM** have been analyzed (see **Figure S.11** and **Table S.4** in the Supporting Information). Line shape of the C*1s* band features three main components for each sample; two at lower energies at 285.0 and c.a. 286.5 eV corresponding to C–N/C=N and C–C/C=C species, respectively;[42] and the third one, assigned to O–C=O carbon atoms,[55] that suffers a shift from 288.9 to 288.1 eV from the bulk to the monolayer. The red-shifted peak is consistent with the presence of carboxylate species in the deposited molecules and is also present in **L·SAM** (see **Table S.4** in the Supporting Information). N*1s* regions show only one component at 399.7 eV assigned to the coordinated nitrogen atoms of the pyridine rings for both samples.[56] Cl*2p* spectrum of **1** features the typical shape of the perchlorate anion centered at 208 eV,[57] while **1·SAM** does not display such signal (**Figure 6**). This is further experimental evidence of the absence of physiosorbed molecules on the surface and of the deprotonation of the carboxylic acid groups upon deposition that compensates the +2 charge of the complexes. This result is in agreement with the conclusion drawn from C*1s* region, MALDI-TOF MS, Raman spectroscopy and with the literature.[49] The Co$2p_{3/2}$ region of the bulk and monolayer display the expected line shape of a Co(II) photoemission spectra (see **Figure 6**).[58,59] In order to get qualitative information about the SCO properties, it was fitted with a similar procedure previously reported for HS/LS-Co(II), where the intensity of the satellites is directly correlated to the paramagnetism, *i.e.* to the spin state of the metal ion.[24] In our approach, the line shapes have been reproduced with five Co$2p_{3/2}$ components (A−E) along with the corresponding Co$2p_{1/2}$ spin−orbit (SO) coupled contributions (A′−E′) weighted by the expected 2:1 ratio (see **Figure 6**). **1** and **1·SAM** Co$2p_{3/2}$ components are composed by a main peak A at 780.4 eV for the bulk and 780.9 eV for the monolayer, integrating about 23.7 and 21.7 % of the overall signal at 170 K and 18.7 and 17.6 % at 300 K, respectively, (for satellite peak positions and contributions see **Table S.5** in the Supporting Information). Else, a sixth component (F) appears at low binding energies for both samples *ca.* 777 eV. We attribute this to the $L_3M_{23}M_{45}$ ($^1P$) Auger peak with a negligible contribution of less than 6%.[60] Semiquantitative analysis and elemental stoichiometric ratios (see **Table S.5** and **Table S.6** in the Supporting Information) are in good agreement with the expected values. This suggests that most of the molecules covering the surface retain the molecular structure found in the bulk.

XPS was also used to get an insight on the electron level population and thus of the spin state. On one hand, experimental results[58,61,62] confirm that SO splitting increases with the number of unpaired 3d-electrons, being closer to 15 eV for LS-Co(II) and 16 eV for HS-Co(II) species, respectively. Indeed, a change in $\Delta_{SO}$ upon spin transition is expected due to different orbital populations in the two spin states, being larger for the HS-Co(II) than for the LS-Co(II). Therefore, the occurrence of SCO can be followed by SO shift.[63] On the other hand, the percentage of the different components is also indicative of the spin state, an increase in the number of unpaired atomic electrons causes an increase in the satellite intensity in XPS.[24,64] Therefore, the ratio of the integrated area ($\Sigma I_{sat}/I Co2p_{3/2}$) is directly related to the intensity of the satellites and has been selected as the most sensitive parameter to follow the spin transition.[58] Co$2p_{3/2}$ XPS region was measured at 170 and 300 K to analyze the differences between low and high temperature species. A $\Delta_{SO}$ of 0.1 has been observed in both samples (from 15.1 to 15.2 in the bulk and from 15.2 to 15.3 in the monolayer) upon increasing the temperature (see **Table S.5**). To confirm the spin transition, together with the high-energy shift of the metal lines, a satellite magnification is also expected. Indeed, the increment in $\Sigma I_{sat}/ICo2p_{3/2}$ values (from 1.70 to 2.42 in **1** and from 1.91 to 2.66 in **1·SAM** at 170 and 300 K, respectively) indicates that the complexes undergo thermal spin transition, confirming the stability of the SCO properties under X-ray irradiation in high vacuum conditions. A higher HS-Co(II) fraction seems to be present in the monolayer, which is in agreement with Raman and confirmed by XAS measurements (see below). It must be taken into consideration that the higher intensity of the main peak in **1·SAM** compare to **1** could be indicative of a small fraction of Co(III) molecules present in the sample, also seen in XAS (see below). Therefore, the higher HS fraction of molecules observed in XPS may be underrated.

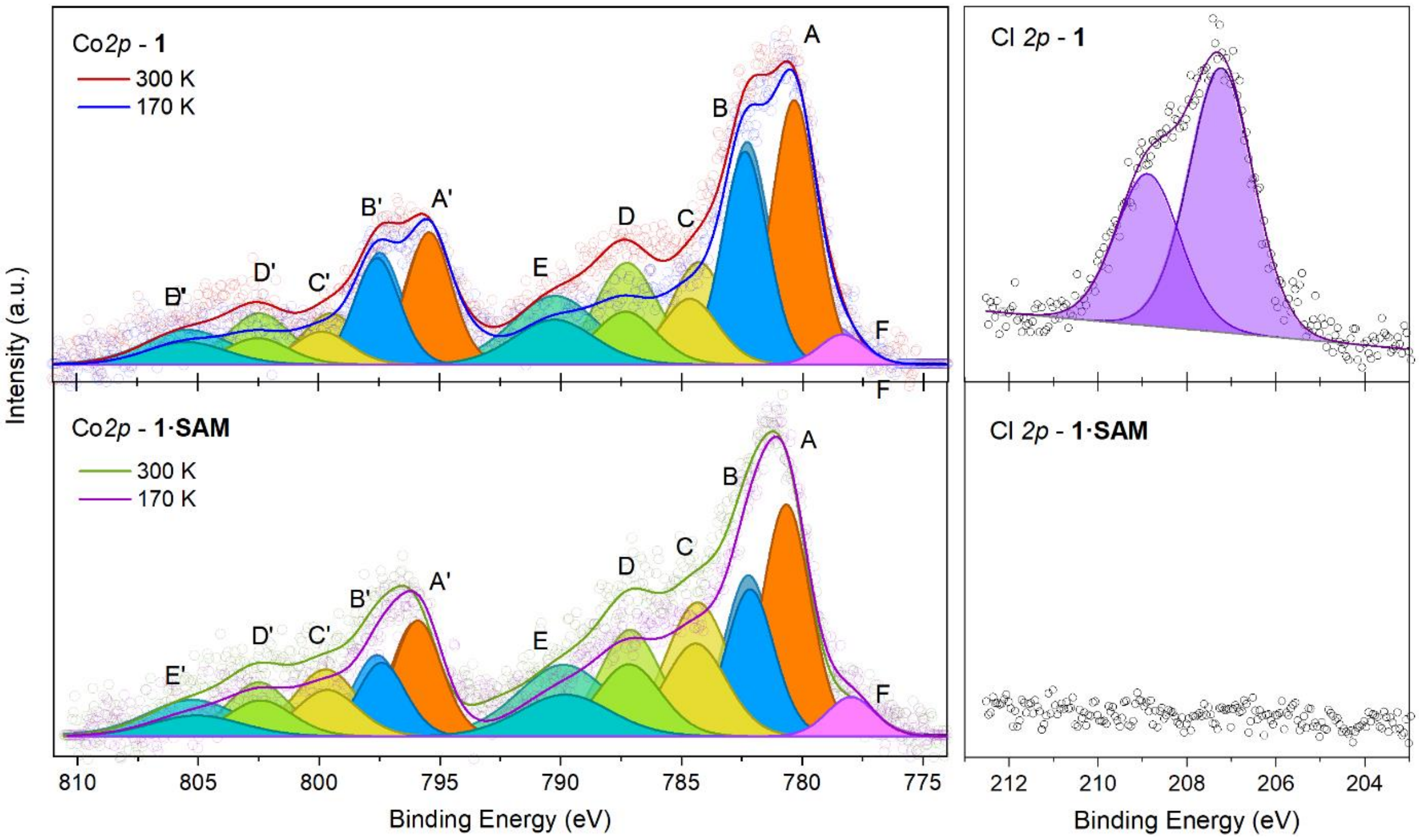

**Figure 6.** Comparison of XPS Co2*p* and Cl2*p* spectra of **1** (top) and **1·SAM** (bottom) along with deconvolution of the peaks and corresponding best fitting lines.

### *2.2.6. X-Ray Absorption Spectroscopy*

XAS spectra were recorded at Co the $L_3$-edge of **1** and **1·SAM** at different temperatures. For the bulk, the spectrum measured at 100K (**Figure S.12a**) which, according to magnetic properties, is consistent with all Co(II) centers in the LS state, is in agreement with the spectra calculated in the Ligand Field Multiplet (LFM) approach for two references of LS-Co(II) in distorted octahedral environment (¡Error! No se encuentra el origen de la referencia.), namely C o(II)-imide[65] and Co(II)-phtalocyanine.[66] The spectrum measured at 370K is not the signature of pure HS Co(II) but is actually a mixture of HS and LS Co(II). A 0.4 fraction of the LS spectrum (i.e., the fraction determined from magnetic measurements) was therefore subtracted, yielding an experimental HS contribution (see **Figure S.12b**) that is consistent with the Co(II) HS theoretical spectrum (¡Error! No se encuentra el origen de la referencia.) a nd with previously published data.[67,68] We note a minor fraction (approx. 3%) of LS octahedral Co(III) (peak at approx. 781.4 eV), a feature that increases both with time and higher photon flux (see **Figure S.14**).

**Figure 7a** shows the temperature dependent XAS spectra of the monolayer in the 125-310K range recorded with a minimized photon flux. Evolution of the signal under soft X-ray irradiation was initially excluded at all temperatures (see **Figure S.15**). However, comparison with the bulk spectra shows a pronounced peak at 781.4 eV arising from Co(III) (see **Figure S.16**). This shows that a fraction of molecules was oxidized during the deposition procedure, which will be quantified below (see **Table S.7**).

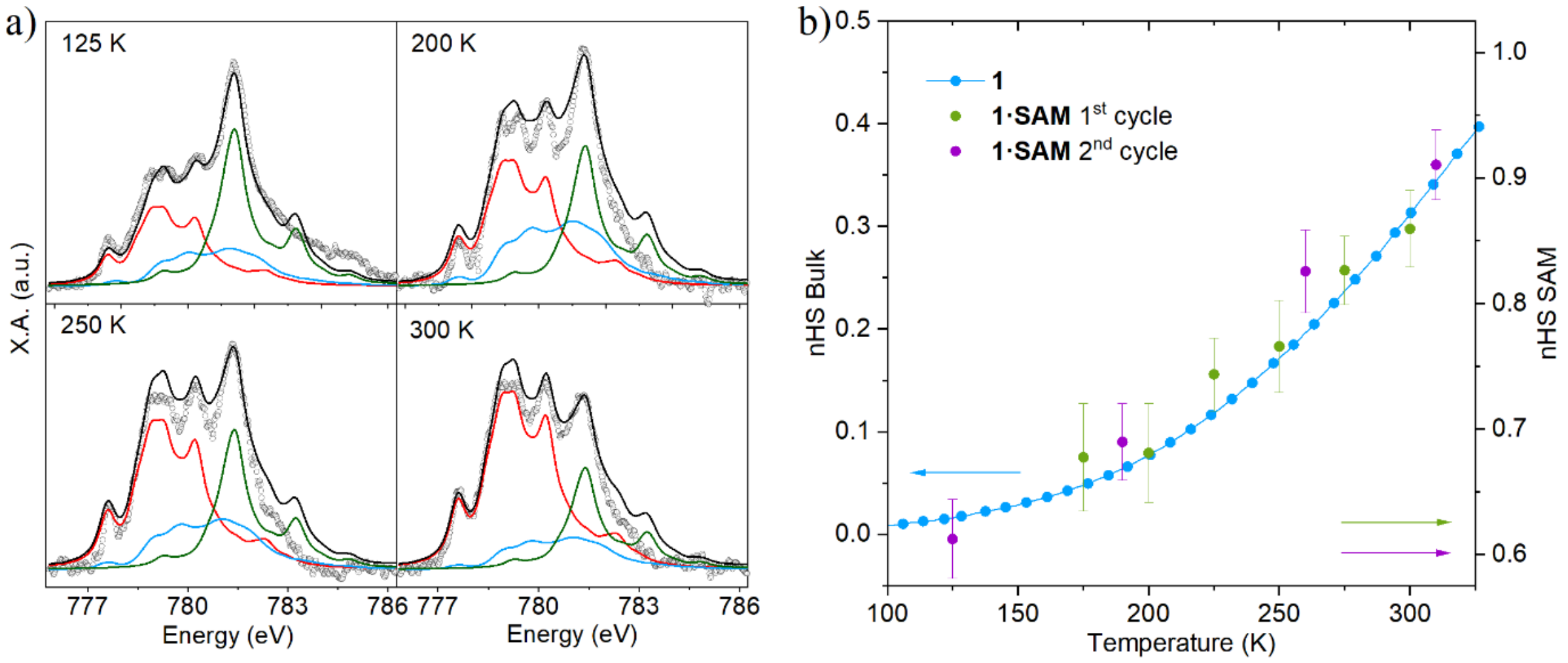

**Figure 7. a)** XAS Co $L_3$-edge of **1·SAM** at each temperature (empty circles) together with HS Co(II) (red curves), LS Co(II) (blue curves) and LS Co(III) (green curve) components. Black line is the best fitting curve with the sum of the three components. **b)** HS Co(II)molar fraction thermal distribution obtained from magnetic measurements of the bulk (blue dots) and from the XAS fitting of **1·SAM** in the first (green dots) and second (purple dots) warming cycles.

To provide a quantitative insight on the temperature-dependence of the Co spin and valence states, the Co $L_3$ XAS spectra of **1·SAM** were fitted using the LFM simulated HS-Co(II) and LS-Co(III) contributions (¡Error! No se encuentra el origen de la referencia.) and the e xperimental LS-Co(II) spectrum of the bulk (**Figure S.12**). The latter was preferred in front of the LFM simulated LS-Co(II) spectrum because the fine structure is strongly dependent on the crystal field distortion parameters used in the calculation, which are currently unknown for **1·SAM**. The results (see **Figure S.17** and **Figure S.18** in the Supporting Information for all the fitted curves) evidence a fraction of oxidized Co(III) molecules, while the rest (Co(II) molecules) undergo a spin transition (see **Figure 7b,** green dots). Two warming cycles were performed to confirm reversibility of the spin transition in the monolayer (**Figure 7b,** purple dots). If we do not consider this Co(III) fraction, and normalize the HS-Co(II) contribution to the total (HS + LS) Co(II) contribution, our results indicate that the monolayer features a higher HS content than the bulk for all the temperatures considered, in good agreement with Raman and XPS experiments. More precisely, the HS fraction increases from *ca.* 61(3) to 91(3) % (see **Table S.8**) for the monolayer and from *ca.* 0 to 30 % for the bulk when warming the sample over the whole temperature range. These results confirm that the thermal spin transition converts approx. 30 % of molecules both in the SAM and in the bulk, but a significant fraction of molecules (*ca.* 60%) remain in the HS state at 125K on the monolayer, which is not the case for the bulk (*ca.* 0%). However, reversibility breaks down when warming **1·SAM** at 370 K. The XAS spectrum at 370K is consistent with the Co(II)-HS state extracted from the bulk (see **Figure S.19a**) but the subsequent cooling does not affect the spin state and the molecules remain blocked in the HS state (see **Figure S.19b).** This result is also in agreement with Raman spectroscopy.

## 3. Conclusion

Herein, we report the synthesis and characterization of a new Co(II) SCO complex based on 4´-(4-carboxyphenyl)-2,2’:6’,2’’-terpyridine ligand. The chemical functionalization of the system provided by the versatile anchoring point and the processability enabled by the higher solubility of the salt, allowed us to prepare SAMs of intact molecules from a diluted solution within a short period of time. AFM, MALDI-TOF MS, Raman spectroscopy and XPS evidences the formation of the deprotonated specie upon surface deposition and confirms the absence of physiosorbed material. Thermal SCO properties of the monolayer have been confirmed with a total of three different techniques: Raman spectroscopy, XPS and XAS pointing to a gradual SCO behavior similar to that observed in the solid state for the protonated complex, which suggests the absence of a cooperative spin transition on the surface as observed previously for monolayers of evaporated SCO complexes. This represents, as far as we are aware, the first example of a SAM of active SCO complexes anchored to a metallic substrate. Finally, our contribution paves the way to the realization of SCO nanostructure-based devices through soft methodologies compared to the high-cost conventional ultra-high vacuum techniques largely employed for this purpose up to now. Further improvements such as a more abrupt SCO and higher stability could be achieved by expanding the aromaticity of the ligand to enhance the intermolecular interactions between the deposited molecules.

**Supporting Information**

Supporting Information is available from the author.
[CCDC 2191823-2191825 contains the supplementary crystallographic data for this paper. These data can be obtained free of charge from The Cambridge Crystallographic Data Centre via www.ccdc.cam.ac.uk/data_request/cif.]

**Acknowledgements**

We acknowledge the financial support from the European Union (COST Action MOLSPIN CA15128), the Spanish MICINN (MAT2017-89993-R, RTI2018-098568-A-I and EQC2018-004888-P co-financed by FEDER and Excellence Unit "María de Maeztu", CEX2019-000919-M), and the Generalitat Valenciana (PO FEDER Program, ref. IDIFEDER/2018/061). Proteomic measurements were performed in the proteomics facility of SCSIE University of Valencia. This proteomics laboratory is a member of Proteored. The authors thank J. M. Martínez, G. Agustí and A. López from the University of Valencia for their help with magnetic measurements, temperature-variable Raman measurements and XAS measurements. XAS experiments were performed at BL29-BOREAS beamline at ALBA Synchrotron with the collaboration of ALBA staff.

**Supporting Information**

# Spin crossover self-assembled monolayer of Co(II) terpyridine derivative functionalized with carboxylic acid groups

*Víctor García-López, Niccolò Giaconi, Lorenzo Poggini, Amelie Juhin, Brunetto Cortigiani, Javier Herrero-Martín, Matteo Mannini,* Miguel Clemente-León* and Eugenio Coronado*

**Experimental Section**

*Synthesis of [Co^II(4´-(4-carboxyphenyl)-2,2':6',2''-terpyridine)$_2$](ClO$_4$)$_2$·4DMA, (**1**):* All chemicals were reagent grade and used as received. A solution of $Co(ClO_4)_2 \cdot xH_2O$ (36.6 mg, 0.10 mmol) in DMA (3 mL) was added dropwise to a solution of HL (70.6 mg, 0.15 mmol) in DMA (3 mL) and the mixture was stirred for 15 minutes. A red-dark solution was formed. Red-dark prismatic crystals of **1** suitable for X-ray diffraction were obtained by slow diffusion of diethyl ether into this solution. Microanalysis shows a Co:Cl ratio of 1:2. Anal. Calcd for $[Co(HL)_2](ClO_4)_2 \cdot 1DMA \cdot 5(H_2O)$ ($C_{48}H_{44}Cl_2CoN_7O_{18}$): C, 50.72; H, 3.90; N, 8.62 %. Found: C, 50.69; H, 3.83; N, 8.48 %.

*Single crystal X-Ray diffraction measurements:* A single crystal of **1** was mounted on a glass fiber loop using a viscous hydrocarbon oil to coat the crystal and then transferred to the cold nitrogen stream for data collection. X-ray data were collected at 120, 300 and 340 K. Higher temperatures leaded to the quick loss of crystallinity of 1. Measurements were performed on a Supernova diffractometer equipped with a graphite-monochromated Enhance (Mo) X-ray Source ($\lambda$ = 0.710 73 Å). The program CrysAlisPro, Oxford Diffraction Ltd., was used for unit cell determinations and data reduction. Empirical absorption correction was performed using spherical harmonics, implemented in the SCALE3 ABSPACK scaling algorithm. The structures were solved with the ShelXT structure solution program[S1] and refined with the SHELXL-2013 program,[S2] using Olex2.[S3] Non-hydrogen atoms were refined anisotropically, and hydrogen atoms were placed in calculated positions refined using idealized geometries (riding model) and assigned fixed isotropic displacement parameters. Crystallographic data of all the compounds are summarized in **Table S2**. Crystallographic data for the structures was deposited in the Cambridge Crystallographic Data Centre, deposition numbers CCDC 2130639-42. These data can be obtained free of charge from The Cambridge Crystallographic Data Centre via www.ccdc.cam.ac.uk/data_request/cif.

*Powder X-ray diffraction (PXRD):* PXRD patterns were performed using a 0.7 mm glass capillary filled with polycrystalline samples of the compounds and mounted and aligned on an Empyrean PANalytical powder diffractometer, using Cu K$\alpha$ radiation ($\lambda$ = 1.541 77 Å). A total of three scans were collected for each compound at room temperature in the 2$\theta$ range of 5-40°.

*Physical measurements:* Co/Cl ratios were measured with a Philips ESEM XL30 scanning electron microscope equipped with an EDAX DX-4 microsonde. Elemental analyses (C, N, and H) were performed with a CE Instruments EA 1110 CHNS Elemental analyzer. Magnetic measurements were performed with a Quantum Design MPMS-XL-5 SQUID magnetometer on powdered polycrystalline samples .

*Substrate and SAM preparation:* Substrates were prepared by thermal evaporation of silver on top of silicon dioxide or glass substrates previously covered with 2-3 nm of chromiun as an adessive layer. Mechanically roughneded silver substrates were prepared by gently apliying pressure with a sharp tip over the surface. They were clean-annealed under an hydrogen flame and washed with EtOH and anhidrous DMA. Then, they were immersed for 6 h in a 1 mM anhydrous DMA solution of the complex. Past that time, they were washed and rinsed several times (6 times) with clean DMA to remove any physiosorbed material and dried under $N_2$ stream. SAMs of **HL** were also prepared for comparison purposes following the same procedure but due to the low solubility of the ligand the solution has to be filtered before immersing the substrates.

*Atomic Force Microscopy (AFM):* AFM measurements were performed with a Digital Instruments Veeco Nanoscope IVa AFM microscope in tapping mode, using silicon tips with a natural resonance frequency of 300 kHz and with an equivalent constant force of 40 N/m.
*Matrix Assisted Laser Desorption Ionization – Time-Of-Flight Mass Spectrometry (MALDI-TOF MS):* The samples were analyzed in a 5800 MALDI TOF/TOF (ABSciex) in positive reflection mode (3000 shots every position; LASER INTENSITY 4200) in a mass range of 200−1500 *m/z*. Previously, the sample and the acquisition method were externally calibrated with a Bruker PSCII solution. In order to do so, 0.5 µL of Bruker PSCII dissolution was spotted on a corner of the sample and allowed to air-dry at room temperature. After this, 0.5 µL of matrix (10mg/mL CHCA (Bruker) in 70% MeCN, 0.1% TFA) was spotted on the previous drop and allowed to air-dry at room temperature. On the diagonal opposite corner 0.5 µL of matrix solution was spotted and allowed to air-dry at room temperature.

*Raman Spectroscopy/Surface-Enhance Raman Spectroscopy:* Spectroscopical characterization was carried out on a LabRAM HR Evolution confocal Raman microscope (Horiba). The measurements were conducted with an excitation wavelength of 532 nm from a Helium Neon Laser source. The laser was focused using a 50× objective (0.8 NA), thus leading to a laser spot with a diameter of *ca.* 300 µm. An exposure time of 3s with 6 accumulations was employed. A CCD camera was used to collect the backscattered light that was dispersed by a 600 grooves per mm grating providing a spectral resolution of ~ 1 $cm^{-1}$. The corresponding Raman spectra were then constructed by processing the data using Lab Spec 5 software. For temperature dependent Raman spectroscopy, a Linkam Scientific THMS600 temperature stage controlled with liquid nitrogen was employed. The measurements were performed after first cooling the sample to the lowest available temperature and then warming it up to the selected temperatures. After reaching the desired temperature, the sample was thermalized for 5 minutes before each measurement. Then, after the first temperature-dependent experiment, the same sample was stored for one week in air and two more cooling and warming cycles were performed. The second cycle was done from the lowest temperature available to 300 K, and the third one until 360 K. Background correction was applied after averaging several scans across the substrate. Then, the intensities were normalized against a temperature-independent feature at 1365 $cm^{-1}$.

*X-ray photoelectron spectroscopy (XPS):* XPS measurements were carried out in a UHV apparatus with a base pressure in the $10^{-10}$ mbar range. Non-monochromatized Al Kα radiation was used for XPS measurements (1486.6 eV, 100 W). The detector was an VSW hemispherical analyzer mounting a 16-channel detector, the angle between the analyzer axis and the X-ray source was 54.44° and the semicone angle of acceptance of the analyzer was 4°. The XPS spectra were measured with a fixed pass energy of 44 eV. The binding energy (BE) scale was calibrated setting the C*1s* signal of the sample at 285 eV.[S7] In order to minimize air exposure and atmospheric contamination, samples were mounted on a sample holder under nitrogen atmosphere. Variable temperature experiments were performed using a liquid nitrogen-based cryostat connected to the XPS sample holder. Every spectrum represented herein results from averaging several spectra collected after 1 h of thermalization at a specific temperature. Stoichiometry was calculated by peak integration, using a theoretically estimated cross-section for each transition.[S8] Semiquantitative analysis has been estimated by areas of the deconvoluted peaks. Components were estimated using CasaXPS, a fitting procedure involving Gaussian–Lorentzian line-shapes was adopted and the background in the spectra was subtracted by means of a linear or Shirley baseline.

*X-Ray Absorption Spectroscopy (XAS):* The experiments were performed at the BL29-BOREAS beamline of the Alba Synchrotron Light Facility (Barcelona, Spain). Co $L_{2,3}$-edge spectra were recorded at different temperatures in HECTOR cryomagnet endstation in the 125 - 370 K range, the chamber pressure being lower than $1 \times 10^{-10}$ mbar, and photon flux $\sim 5 \times 10^{11}$ $s^{-1}$ in total electron yield (TEY) method with an energy resolution of ~70 meV. The spectra were recorded using alternatively left and right circularly polarized X-rays produced by an APPLE II undulator and a superconducting split coil setup generating a magnetic field up to 6 T in the direction of propagation of the incident photons.

## Theoretical Section

*Ligand Field Multiplet (LFM) calculations:*

In order to identify and quantify the spin and valence states of Co in the T-dependent Co $L_3$-edge XAS spectra, reference spectra for Co(II)-HS, Co(III)-LS and Co(II)-LS were calculated in the LFM using the Quanty code[S9] in the 125-370K temperature range. The Co(III) spectrum was simulated in $O_h$ symmetry with 10Dq = 2.0 eV, $\zeta_{3d}$ = 0.074 eV and $\kappa$ = 0.7 and the Co(II)-HS spectrum was simulated in $D_{4h}$ symmetry with 10Dq = 1.0 eV, $\zeta_{3d}$ = 0.022 eV, Ds = 0.05 eV, Dt = 0 and $\kappa$ = 0.7, similar to the parameters used in Ref. S7. For Co(II)-LS, calculations were performed in $D_{4h}$ symmetry with $\kappa$ = 0.7, using crystal field parameters of (10Dq = 2.11 eV, Ds = 0.16 eV, Dt = 0.08 eV) and (10Dq = 2.3 eV, Ds = 0.5 eV, Dt = 0.2 eV) close to those determined for the Co(II)-LS state of Co-imide of Ref. S8 and for the Co(II)-LS phtalocyanine of Ref. S9 respectively. Isotropic spectra were calculated and plotted with a Lorentzian broadening of 0.4 eV (FWHM for the $L_3$ edge) and 0.8 eV (FHWM for the $L_2$ edge) and a Gaussian broadening of 0.15 eV (FWHM). Calculated XAS intensities were normalized with respect to the sum rule on the number of holes[S10], with n = 6 for Co(III) and n = 7 for Co(II) (**Figure S13**).

*Characterization of [Co$^{II}$(HL)$_2$](ClO$_4$)$_2$·4DMA (L = 4´-(4-carboxyphenyl)-2,2’:6’,2’’-terpyridine), (**1**)*

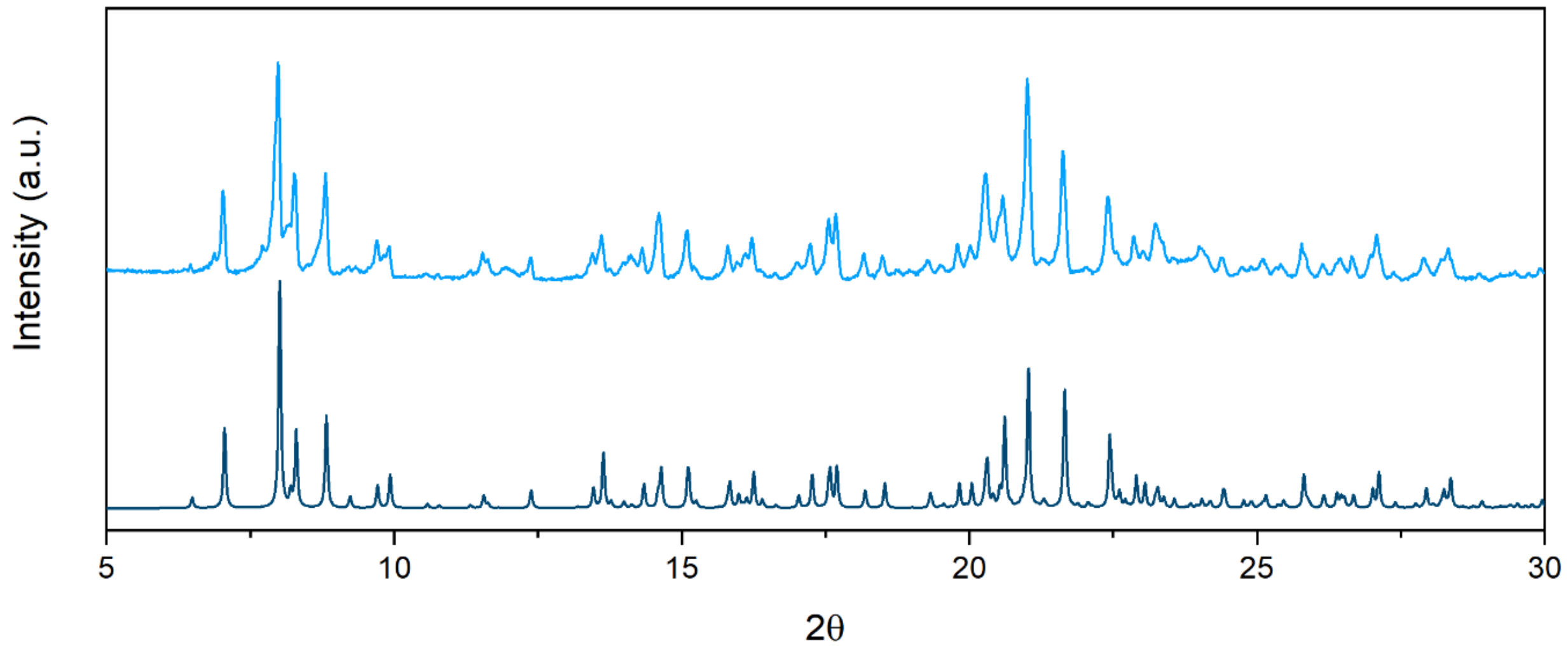

**Figure S.1.** Experimental PXRD pattern (blue) and simulated one from the structure solved at 300 K (dark blue) of **1.**

Elemental analysis results of **1** indicate that two solvent molecules present in the structure are replaced by five water molecules in contact with ambient air. For this reason, experimental PXRD presents small differences with respect to the simulated one from the structure solved under nitrogen stream.

*Structure of [Co$^{II}$(HL)$_2$](ClO$_4$)$_2$·4DMA, (**1**)*

O atoms of one $ClO_4^-$ are disordered. They have been modelled as six O atoms for 120 K and eight O atoms for 300 K and 340 K with 50% occupancy. At 300 K and 340 K, two DMA solvent molecules are disordered as well. They have been solved as four molecules with an occupancy of 50% each.

**Table S.1.** Co-N bond lengths for **1** at 120K, 300K and 340K.

| **120K** | | | **300K** | | | **340K** | | |
|---|---|---|---|---|---|---|---|---|
| **Bond** | | **Length (Å)** | **Bond** | | **Length (Å)** | **Bond** | | **Length (Å)** |
| Co1 | N1 | 2.016(2) | Co1 | N1 | 2.085(3) | Co1 | N1 | 2.097(4) |
| Co1 | N2 | 1.872(2) | Co1 | N2 | 1.950(3) | Co1 | N2 | 1.969(3) |
| Co1 | N3 | 2.011(2) | Co1 | N3 | 2.095(3) | Co1 | N3 | 2.108(4) |
| Co1 | N4 | 2.162(2) | Co1 | N4 | 2.140(3) | Co1 | N4 | 2.147(4) |
| Co1 | N5 | 1.928(2) | Co1 | N5 | 1.954(3) | Co1 | N5 | 1.965(4) |
| Co1 | N6 | 2.139(2) | Co1 | N6 | 2.116(3) | Co1 | N6 | 2.125(4) |

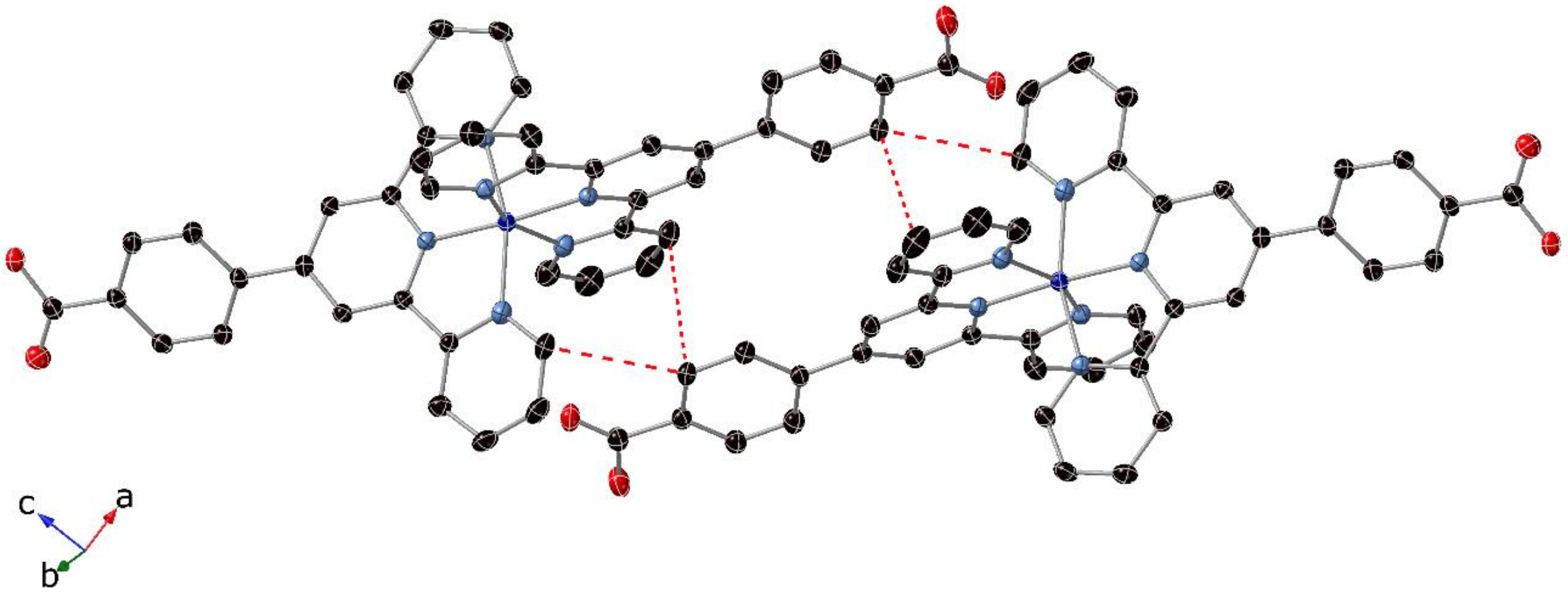

**Figure S.2.** View of the supramolecular $\pi\cdots\pi$ and CH$\cdots\pi$ interactions between pairs of molecules of **1** (C (black), N (blue), O (red), Co (dark blue), counter ions, solvent molecules, and hydrogen atoms are omitted for clarity, $\pi\cdots\pi$ interactions are represented with dashed red lines,).

**Table S.2.** Crystal data and structure refinement for **1**.

| Compound reference | **1** | **1** | **1** |
|---|---|---|---|
| Empirical formula | $C_{60}H_{66}Cl_2CoN_{10}O_{16}$ | $C_{60}H_{66}Cl_2CoN_{10}O_{16}$ | $C_{60}H_{66}Cl_2CoN_{10}O_{16}$ |
| Formula weight | 1313.05 | 1313.05 | 1313.05 |
| Temperature/K | 120.00(10) | 299.4(10) | 340(4) |
| Crystal system | triclinic | triclinic | triclinic |
| Space group | P-1 | P-1 | P-1 |
| a/Å | 12.3368(4) | 12.5846(5) | 12.6608(6) |
| b/Å | 13.5063(5) | 13.6472(4) | 13.6988(5) |
| c/Å | 18.0712(5) | 18.2438(6) | 18.2866(5) |
| α/° | 90.221(2) | 91.119(3) | 91.312(3) |
| β/° | 94.068(2) | 93.664(3) | 93.524(3) |
| γ/° | 94.159(3) | 94.085(3) | 94.057(3) |
| Volume/Å$^3$ | 2995.49(17) | 3118.03(19) | 3156.5(2) |
| Z | 2 | 2 | 2 |
| $\rho_{calc}$g/cm$^3$ | 1.456 | 1.399 | 1.382 |
| μ/mm$^{-1}$ | 0.455 | 0.437 | 0.432 |
| F(000) | 1370.0 | 1370.0 | 1370.0 |
| Crystal size/mm$^3$ | 0.45 × 0.35 × 0.2 | 0.45 × 0.35 × 0.2 | 0.45 × 0.35 × 0.2 |
| Radiation | MoKα (λ = 0.71073) | MoKα (λ = 0.71073) | MoKα (λ = 0.71073) |
| 2Θ range for data collection/° | 6.478 to 60.124 | 6.604 to 58.24 | 6.426 to 58.3 |
| Index ranges | $-16 \le h \le 16$, $-18 \le k \le 18$, $-25 \le l \le 24$ | $-17 \le h \le 16$, $-17 \le k \le 18$, $-23 \le l \le 24$ | $-17 \le h \le 17$, $-17 \le k \le 18$, $-23 \le l \le 24$ |
| Reflections collected | 48487 | 48421 | 48772 |
| Independent reflections | 15595 [$R_{int}$ = 0.0504, $R_{sigma}$ = 0.0628] | 14990 [$R_{int}$ = 0.0676, $R_{sigma}$ = 0.1082] | 15181 [$R_{int}$ = 0.0708, $R_{sigma}$ = 0.1160] |
| Data/restraints/parameters | 15595/31/840 | 14990/59/960 | 15181/53/840 |
| Goodness-of-fit on F$^2$ | 1.056 | 1.034 | 1.039 |
| Final R indexes [I>=2σ (I)] | $R_1$ = 0.0666, $wR_2$ = 0.1621 | R1 = 0.0698, wR2 = 0.1415 | $R_1$ = 0.0816, $wR_2$ = 0.1853 |
| Final R indexes [all data] | $R_1$ = 0.0911, $wR_2$ = 0.1831 | R1 = 0.1699, wR2 = 0.2014 | $R_1$ = 0.1988, $wR_2$ = 0.2698 |
| Largest diff. peak/hole / e Å$^{-3}$ | 2.06/-0.73 | 0.40/-0.40 | 0.53/-0.45 |

*Magnetic properties of [Co(L)$_2$]·5H$_2$O*

Temperature dependence of $\chi_M$T of [Co(L)$_2$]·5H$_2$O shows values close to 0.5 cm$^3$·K·mol$^{-1}$ from 5 to 250 K typical of LS Co(II). At higher temperature there is a gradual increase to reach a maximum at 360 K of 1.21 cm$^3$·K·mol$^{-1}$. Interestingly, at higher temperatures there is a decrease of $\chi_M$T to reach a minimum at 374 K (0.75 cm$^3$·K·mol$^{-1}$). This is a clear indication that a reverse spin transition takes place in this temperature range. At higher temperatures there is an abrupt increase of $\chi_M$T to reach values close to 2.2 at 388 K. On lowering the temperature, the abrupt spin transition is shifted to lower temperatures with a thermal hysteresis of 30 K. Again, there is a partial reverse spin transition from 340 to 300 K. Reverse spin transition in Co(II) complexes have been reported and have been related to structural phase transitions.[S11] In this case, preliminary variable-temperature PXRD experiments suggest the presence of several phases; the initial hydrate one (**phase I**), which is lost above 350 K; the dehydrated **phase III**, which present an abrupt spin transition from 340 to 400 K in cooling and heating modes; and the intermediate **phase II** at around 310 K in the first heating and cooling cycle undergoing a gradual spin transition. The structural phase transitions from **phases II and III** could explain the reverse spin transition.

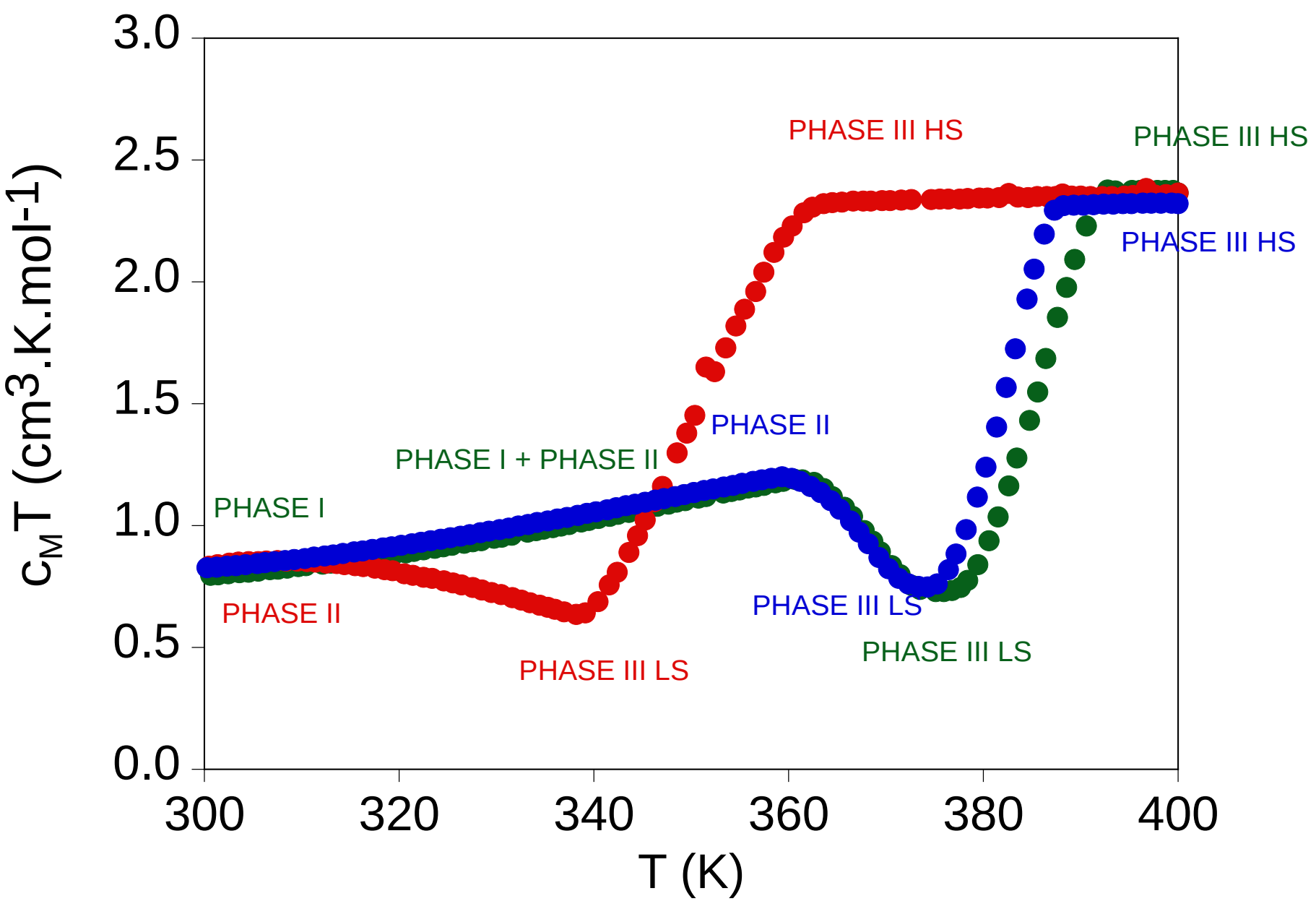

**Figure S.3.** Temperature dependence of the product of the molar magnetic susceptibility and the temperature ($\chi_M$T) of Co(L$_2$) (green, first heating, red, first cooling and blue, second heating). Scan rates 2 K·min$^{-1}$ for the first heating and cooling and 1 K·min$^{-1}$ for the second heating.

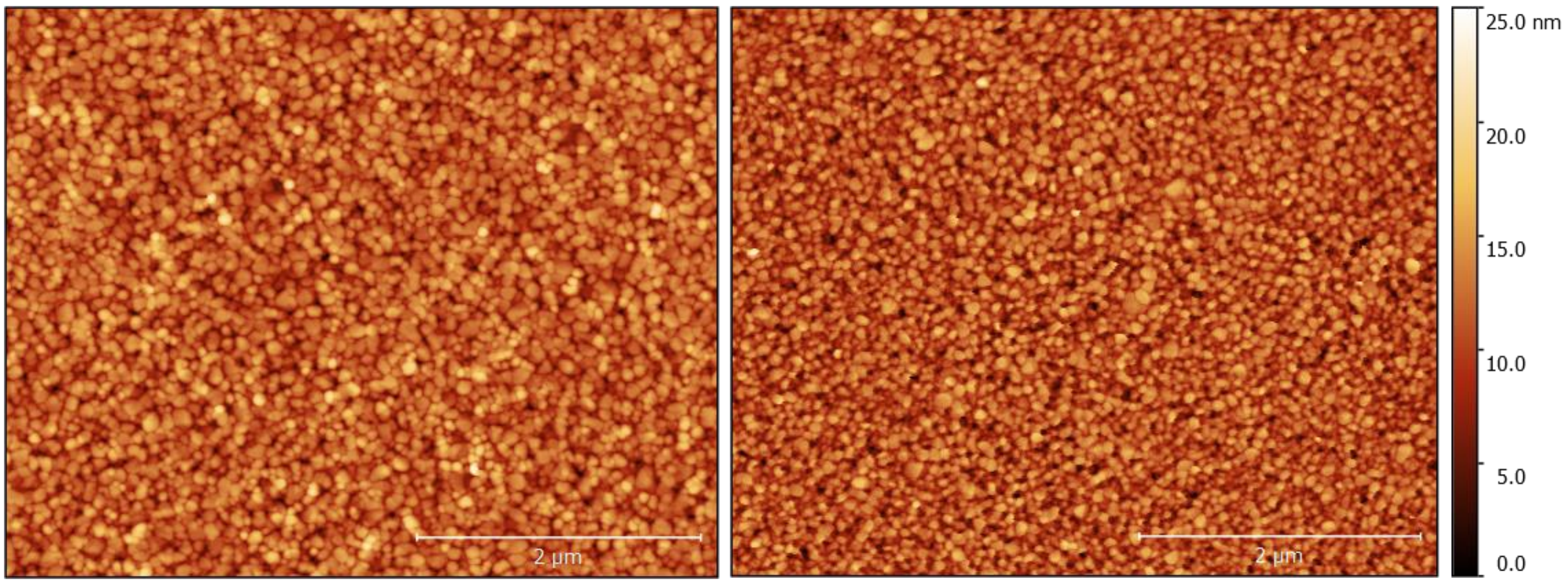

**Figure S.4.** AFM topographic images of silver substrates of **1·SAM** (left) and **L·SAM** (right). The scale bar represents 2µm.

**Table S.3.** MALDI-TOF peaks expected and experimentally found for **1** and **1·SAM**.

| **Fragment** | **Theor.** *(m/z)* | **1** *(m/z)* | **1·SAM** *(m/z)* |
|---|---|---|---|
| $[Co(LH)_2]^+$ | 764.17 | 764.16 | 764.15 |
| $[Co(LH+H+Na)(L+Ag)]^+$ | 895.06 | n/d | 895.01 |
| $[LH+H]^+$ | 354.12 | 354.11 | 354.12 |
| $[L+Ag+H]^+$ | 460.02 | n/d | 460.03 |
| $[Co(LH)]^+$ | 412.05 | 412.00 | 412.04 |
| $[Co(L)(CHCA^{a)})]^+$ | 600.07 | 600.08 | 600.07 |
| $[Co(L)(CHCA)(CHCA\text{-}H+Na)]^+$ | 811.11 | n/d | 811.09 |
| $[Co(LH\text{-}COOH+H)(L+Ag)]^+$ | 827.07 | n/d | 827.07 |
| $[Co(LH\text{-}COOH+2H+Na)(L+Ag)]^+$ | 851.07 | n/d | 851.03 |
| $[Co(LH)(ClO_4)]^+$ | 511.0 | 510.99 | n/d |

a) CHCA is abbreviation for α-cyano-4-hydroxycinnamic acid and is present in the matrix used for MALDI-TOF MS measurements (see Experimental Section).

*Raman discussion*

The most characteristic feature of terpy systems is the asymmetric breathing vibration of pyridine rings located at *ca.* 991 $cm^{-1}$. It is known to be independent of the 4-substituted position but not of metal coordination.[S12] Therefore, the shift observed in the breathing mode above 1020 $cm^{-1}$ in the bulk and monolayer samples of **1** with respect to the monolayer of the ligand was expected. Such shift is also observed in the Raman spectra of $[Co(L)_2]\cdot 5H_2O$[S13] in the solid state measured as reference (see **Figure 4**). No big differences are seen in the 200 – 900 $cm^{-1}$ region except for the carboxylic/carboxylate bands already mentioned in the manuscript. Based on the literature, the stretching vibration region is known to be mainly governed by coordinative terpy modes such as ring-stretching and in plane deformation modes,[S14,S15] although a small contribution of modes coming from the carboxylic/carboxylate groups is also expected.[S16] Nevertheless, we were able to identify the peak at *ca.* 1365 $cm^{-1}$ assigned to the 4-substituted position of the pyridine ring which has been taken as a weak temperature-dependent mode.[S12]

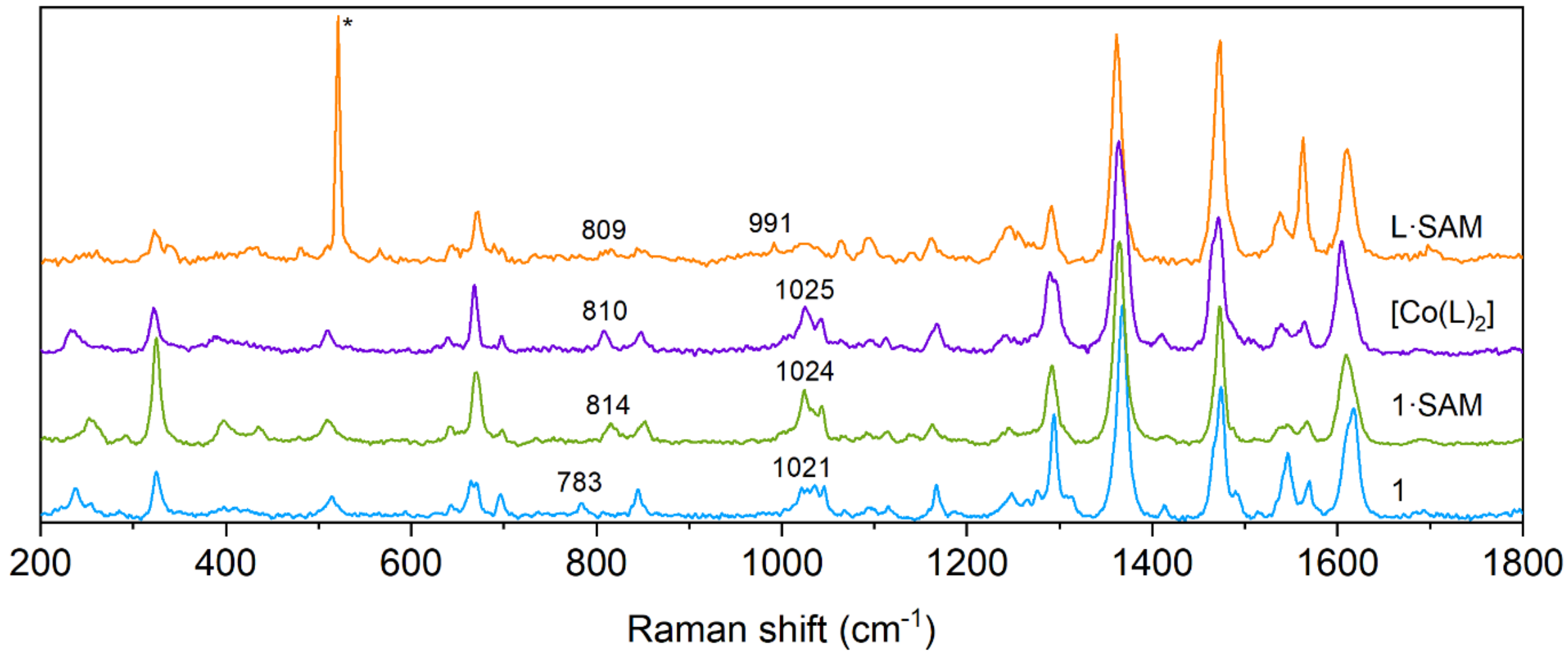

**Figure S.5.** Raman fingerprints of **1** (blue), **1·SAM** (green), **$[Co(L)_2]\cdot 5H_2O$** (purple) and **L·SAM** (orange) in the 200-1800 $cm^{-1}$ at 100K. Peaks marked with asterisk are due to the exposed silicon dioxide after the preparation of the mechanically roughened surfaces.

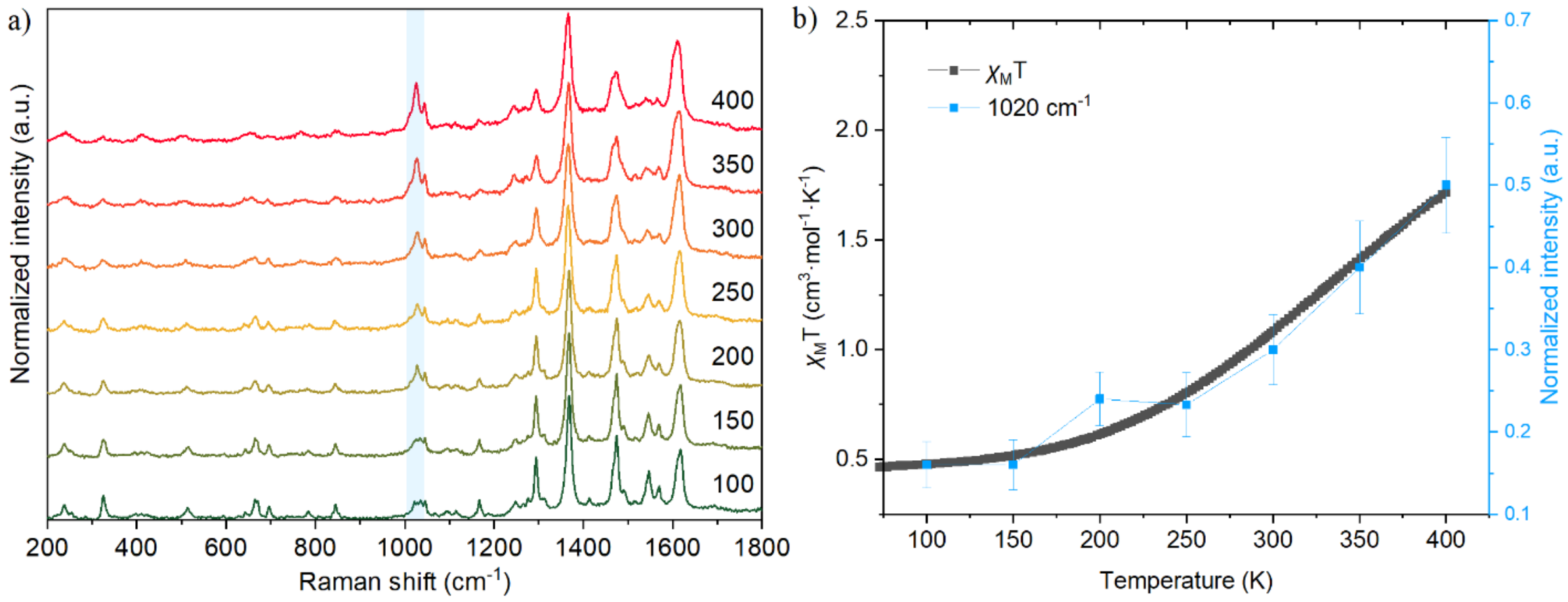

**Figure S.6. a)** Temperature-dependent Raman spectra of **1**. Highlighted in blue the pyridine ring breathing mode at 1020 cm$^{-1}$. **b)** $\chi_M T$ vs temperature (black) and normalized intensity value of the feature at 1020 cm$^{-1}$ (blue squares) for **1**. Error bars have been calculated from the median absolute deviation.

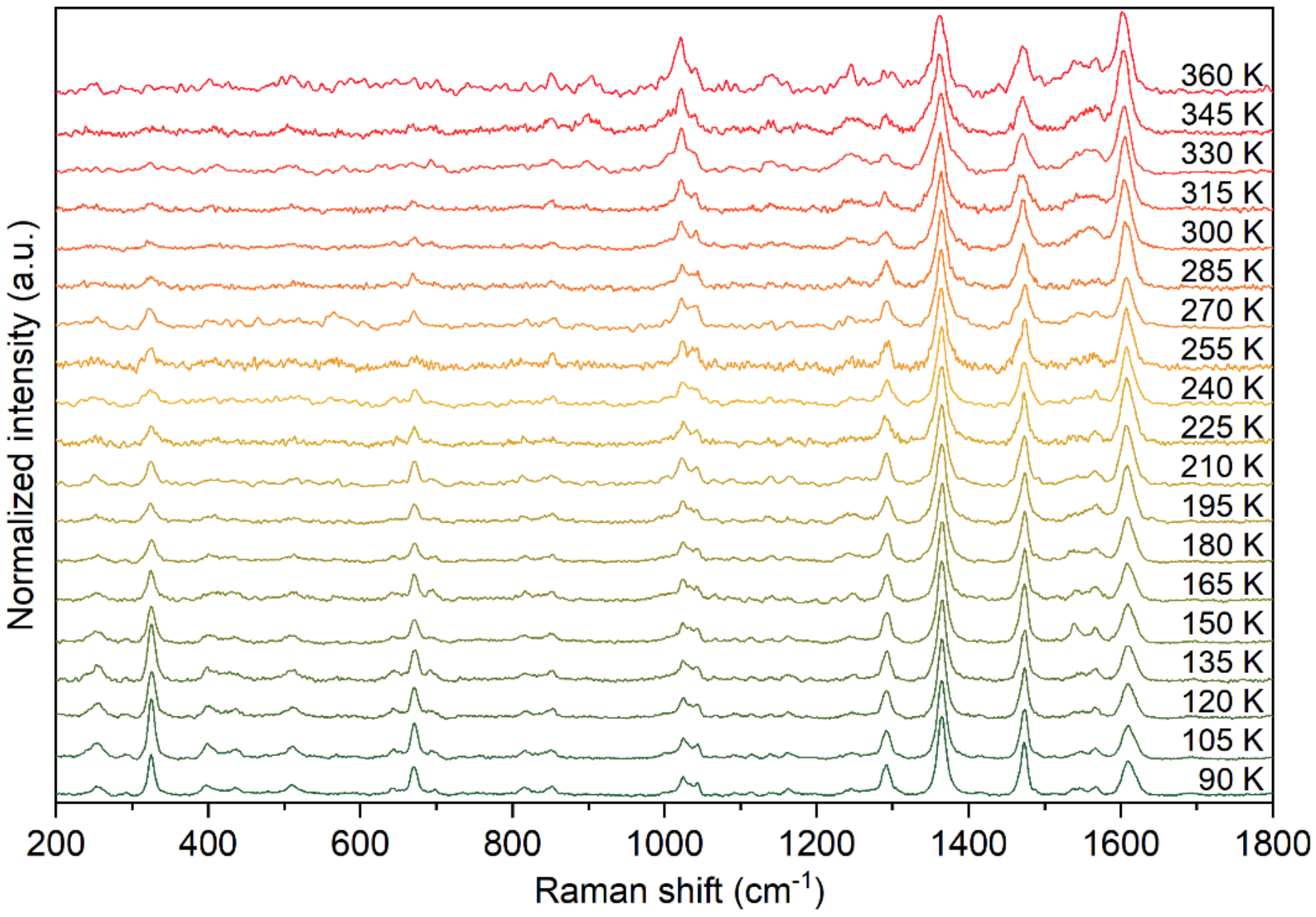

**Figure S.7.** Temperature dependent Raman spectra of **1·SAM** in the 200-1800 cm$^{-1}$ range from 90 to 360 K.

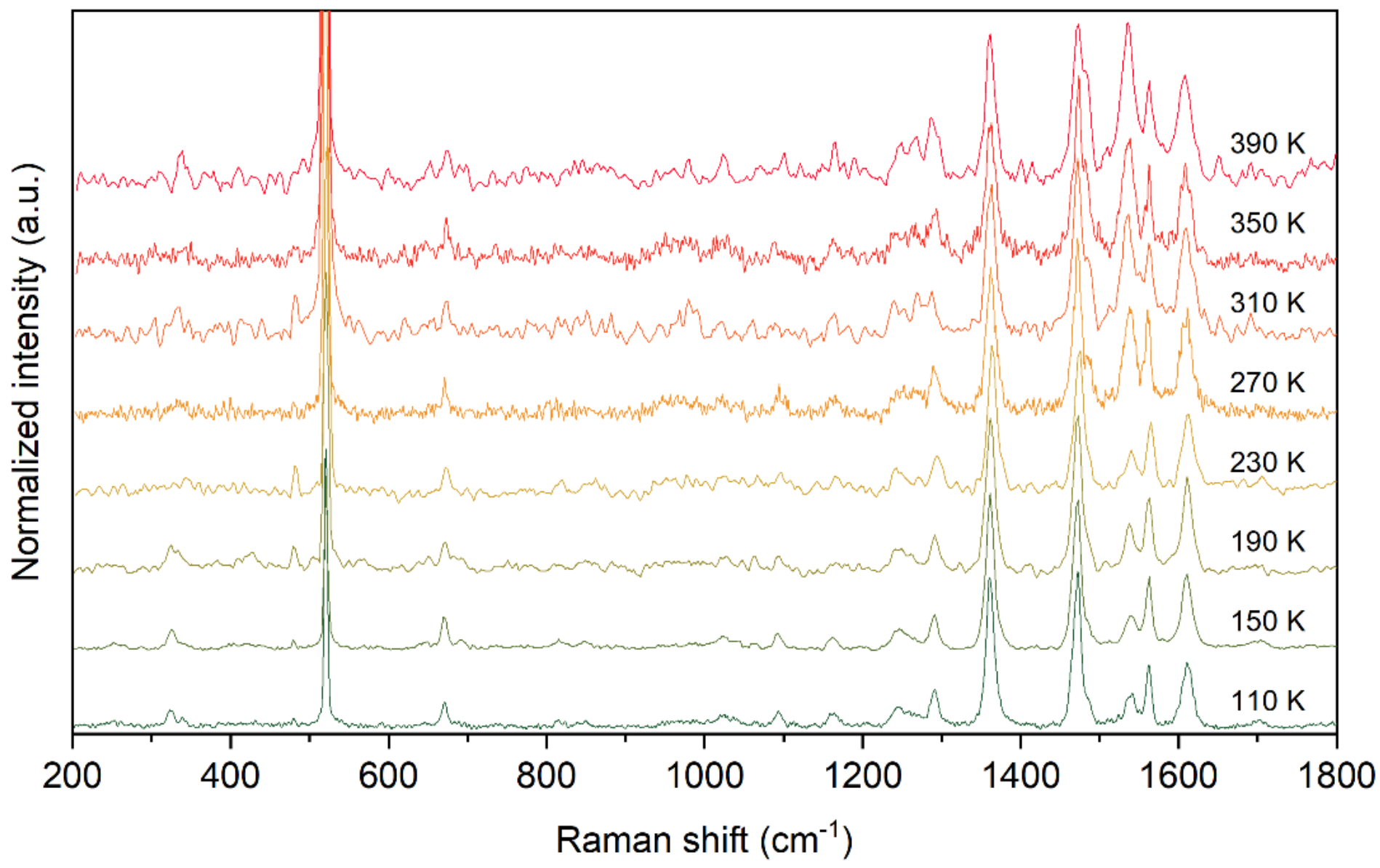

**Figure S.8.** Temperature dependent Raman spectra of **L·SAM**.

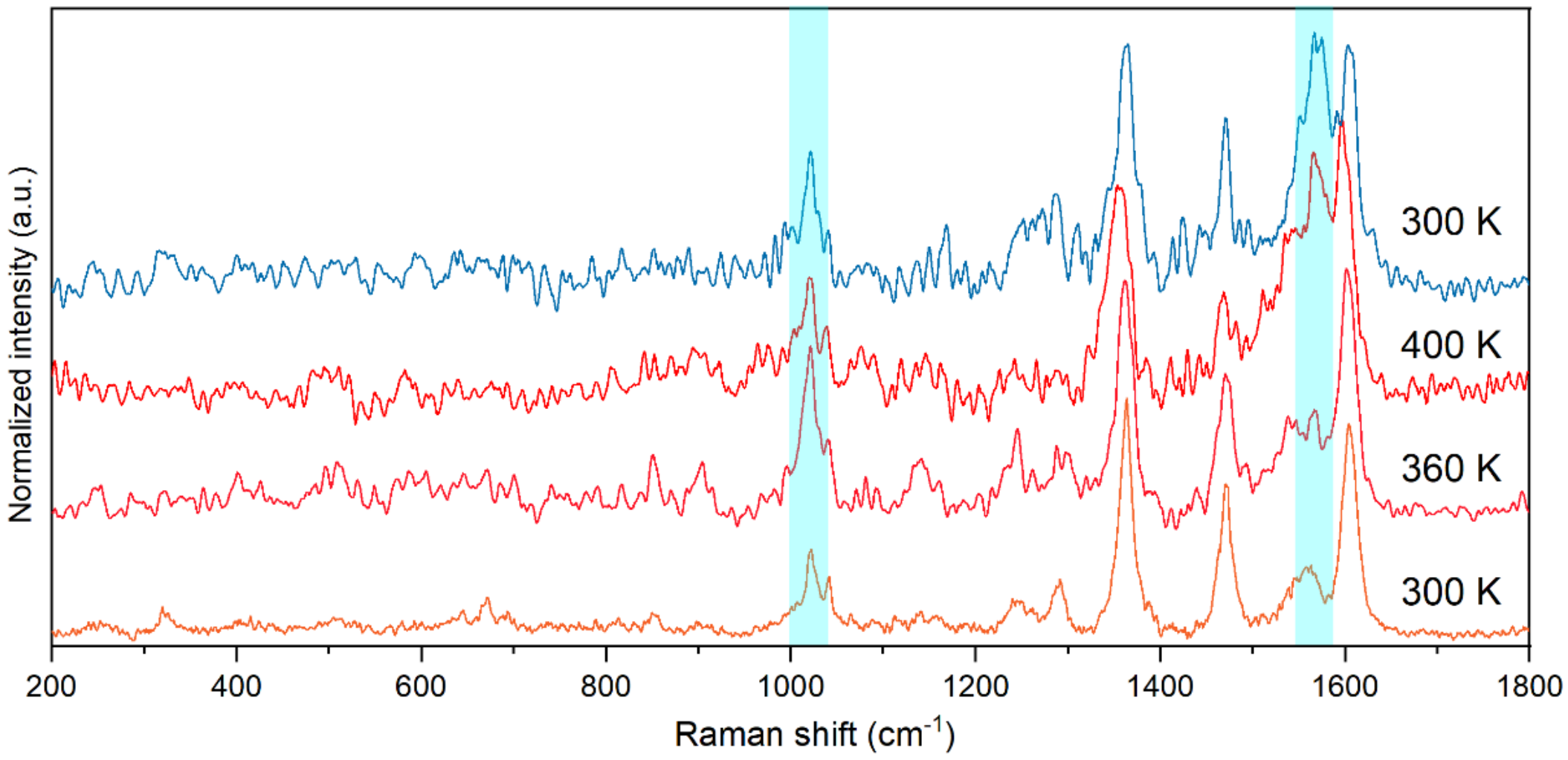

**Figure S.9.** Raman spectra of **1·SAM** before and after heating the sample at 400 K. Irreversible changes in the spectra at 1567 $cm^{-1}$ and 1021 $cm^{-1}$ (highlighted in cyan) take place.

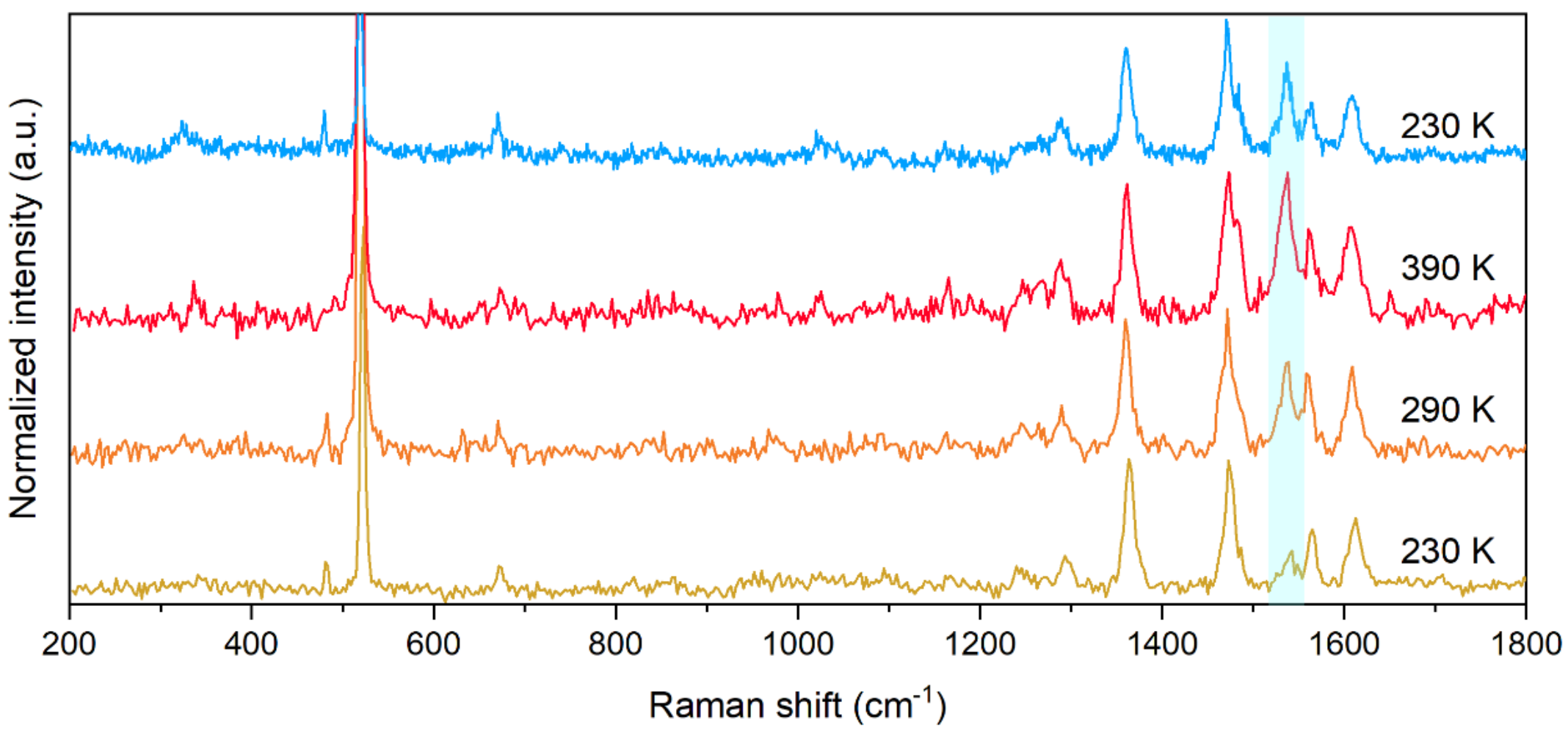

**Figure S.10.** Raman spectra of **L·SAM** before and after heating the sample at 390 K. Irreversible changes in the spectra at *ca.* 1550 $cm^{-1}$ (highlighted in cyan) take place.

Irreversible changes may be related to desorption and/or reorganization of the SAM leading to new conformations on the surface, as it has been observed with this type of molecules.[S17,S18] We think that **1·SAM** also suffers from this conformational effect but, due to the formation of the cobalt complex and/or a different self-assembled structure on the surface, it is not effective until higher temperatures compared to **L·SAM** (see above). This is in agreement with the blocking of the Co(II) HS state obtained in XAS after heating the sample.

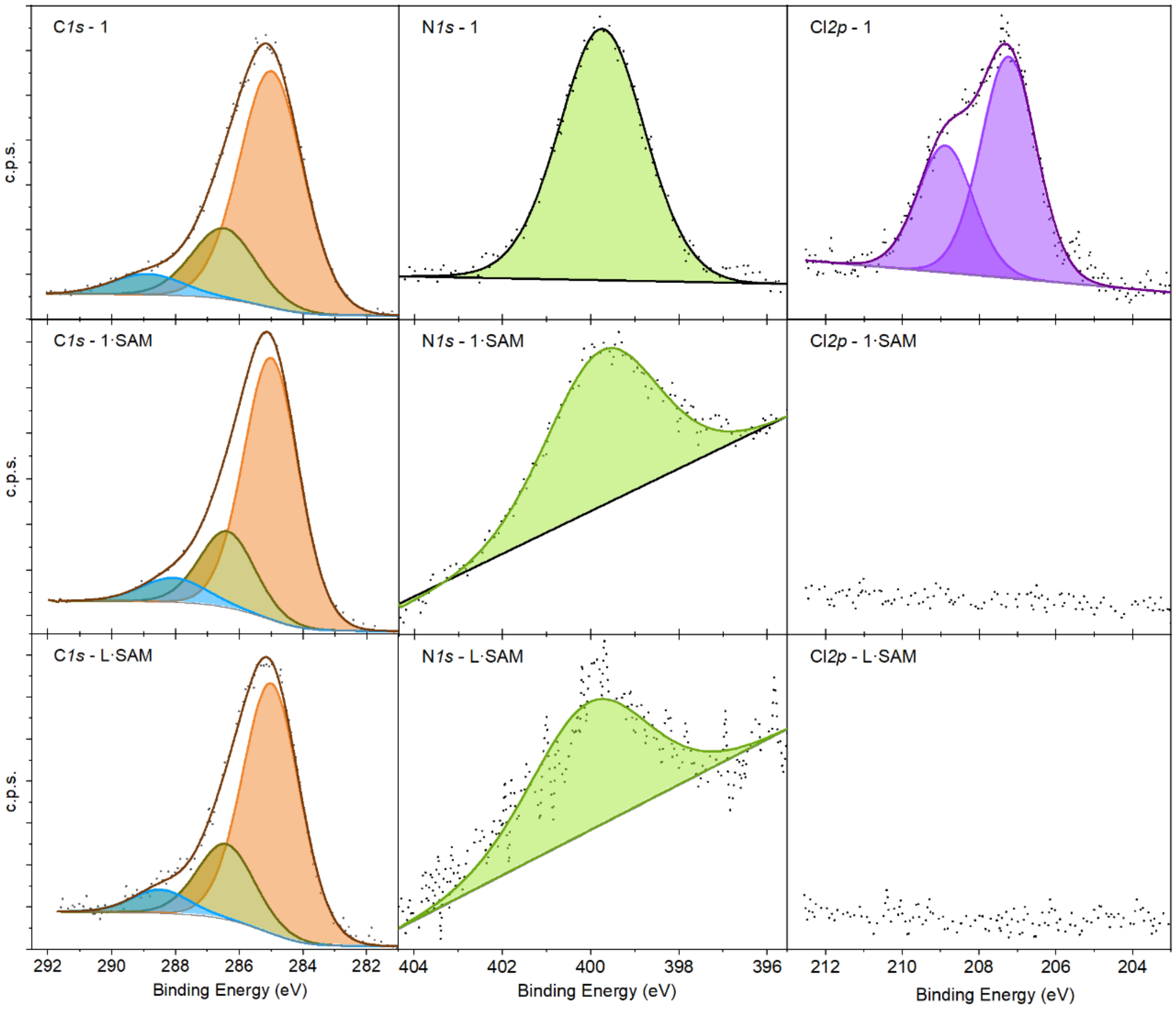

**Figure S.11.** C*1s*, N*1s* and Cl*2p* XPS regions of **1**, **1·SAM** and **L·SAM** along with the best fitting model.

**Table S.4.** XPS peak position and estimated and theoretical percentages (relative error about 5%) of C*1s* and N*1s* elements for compound **1**, **1·SAM** and **L·SAM**.

| Element | **1** | **1·SAM** | **L·SAM** | Theory |
|---|---|---|---|---|
| | B. E. (eV) - % | B. E. (eV) - % | B. E. (eV) - % | % |
| C | 285.0 - **71.6** | 285.0 **71.4** | 284.9 **71.8** | **73.3** |
| | 286.5 **21.4** | 286.4 **20.8** | 286.3 **21.7** | **20.0** |
| | 288.9 **7.0** | 288.1 **7.7** | 288.3 **6.7** | **6.7** |
| N | 399.7 **100** | 399.7 **100** | 399.9 **100** | **100** |
| Cl | 207.2/208.9 **100** | - | - | - |

**Table S.5.** Peak position of the spectral components employed for least squares fitting of the Co2p XPS binding energies (B.E.) for **1** and **1·SAM** at 170 K and 300 K. Integrated areas are reported in percentages for each component in brackets. Spin–orbit splitting ($\Delta_{SO}$) and intensity ratios of the satellites ($\Sigma I_{sat}/ICo2p_{3/2}$) are also gathered.

| | Co 2p | | | | | | |
|---|---|---|---|---|---|---|---|
| | A | B | C | D | E | $\Delta_{SO}$ | $I_{sat}/ICo2p_{3/2}$ |
| | B.E. (%) | B.E. (%) | B.E. (%) | B.E. (%) | B.E. (%) | eV | |
| **1** (170 K) | 780.4 **(23.7)** | 782.4 **(19.1)** | 784.7 **(7.4)** | 787.3 **(6.6)** | 790.3 **(7.2)** | 15.1 | 1.70 |
| **1** (300 K) | 780.4 **(18.7)** | 782.3 **(16.3)** | 784.3 (9.2) | 787.2 **(10.3)** | 790.2 **(9.5)** | 15.2 | 2.42 |
| **1·SAM** (170 K) | 780.7 **(21.7)** | 782.2 **(13.8)** | 784.4 **(11.2)** | 787.2 **(9.0)** | 789.8 **(7.4)** | 15.3 | 1.91 |
| **1·SAM** (300 K) | 780.9 **(17.6)** | 782.4 **(12.4)** | 784.6 **(13.6)** | 787.3 **(10.4)** | 790.1 **(10.3)** | 15.4 | 2.66 |

**Table S.6.** Theoretical and XPS estimated atomic semiquantitative analysis and ratios for bulk and monolayer of **1**.

| | Semiquantitative analysis | | | Element ratios | | | Carbon ratios | | |
|---|---|---|---|---|---|---|---|---|---|
| | Co | N | C | Co/Co | Co/N | Co/C | Co/C-C | Co/C-N | $Co/COO^-$ |
| Theoretical | 2.0% | 11.8% | 86.3% | 1 | 0.17 | 0.023 | 0.03 | 0.09 | 0.27 |
| **1** | 1.6% | 6.4% | 92.1%[a] | 1 | 0.24 | 0.017 | 0.02 | 0.07 | 0.22 |
| **1·SAM** | 2.0% | 11.4% | 86.6% | 1 | 0.18 | 0.024 | 0.03 | 0.10 | 0.26 |

a) Although a homogeneous coverage of all the surface was done spreading a thick layer of powder, the higher value of the carbon semiquantitative analysis of **1** is probably coming from the carbon tape used to attach the powder to the sample holder.

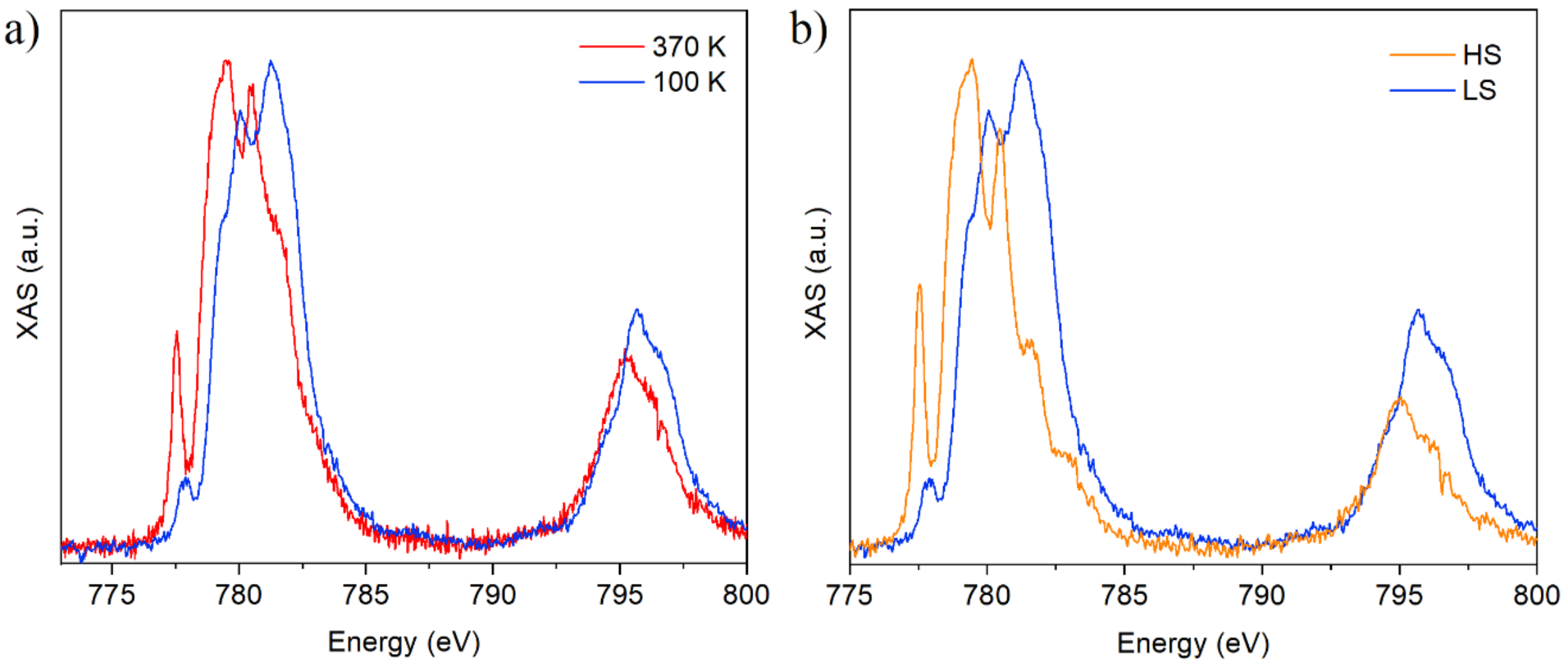

**Figure S.12.** XAS Co $L_{3,2}$-edge of the desolvated bulk **a)** at the highest available temperature (370 K) and 100 K and **b)** LS and HS reference spectra. The HS spectrum was obtained from subtracting a 0.4 fraction of the experimental LS spectrum.

The XAS spectrum of the Co $L_3$-edge at 100 K for the bulk is consistent with the Co(II) LS theoretical spectra calculated in the LFM approach, especially the with Co-imide (**Figure S.13**). Thus, in order to get the HS spectra of our complex, a 0.4 fraction of the low temperature experimental spectrum (LS state), according to magnetic measurements, was subtracted from the 370 K XAS spectrum of the bulk. The obtained spectrum is in good agreement with the theoretical one (red line in **Figure S.13**).

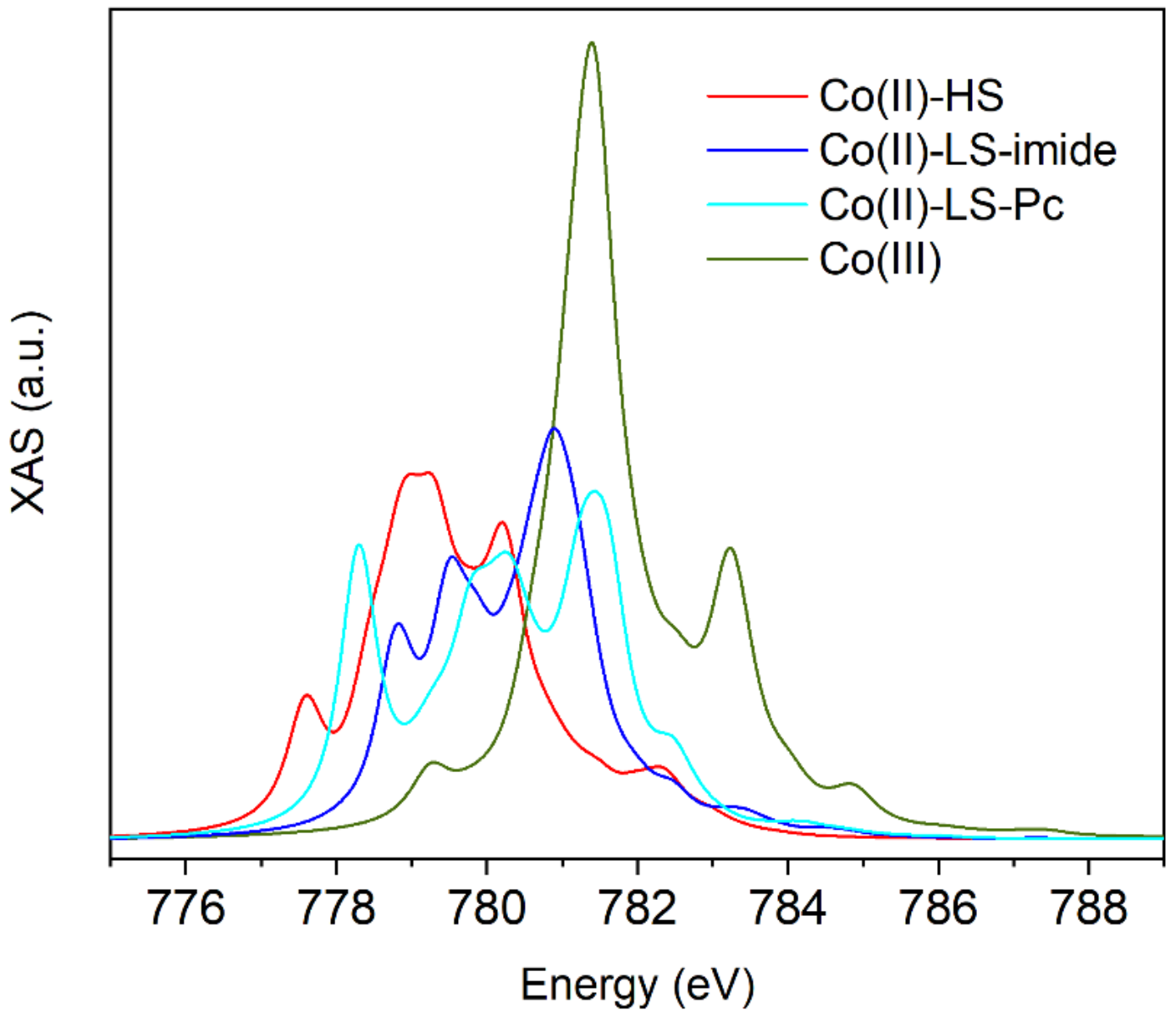

**Figure S.13.** Co $L_3$-edge XAS spectra calculated in the LFM approach for Co(II) LS (at T = 125K), Co(II) HS (at T = 370K) and Co(III) (at T = 125K).

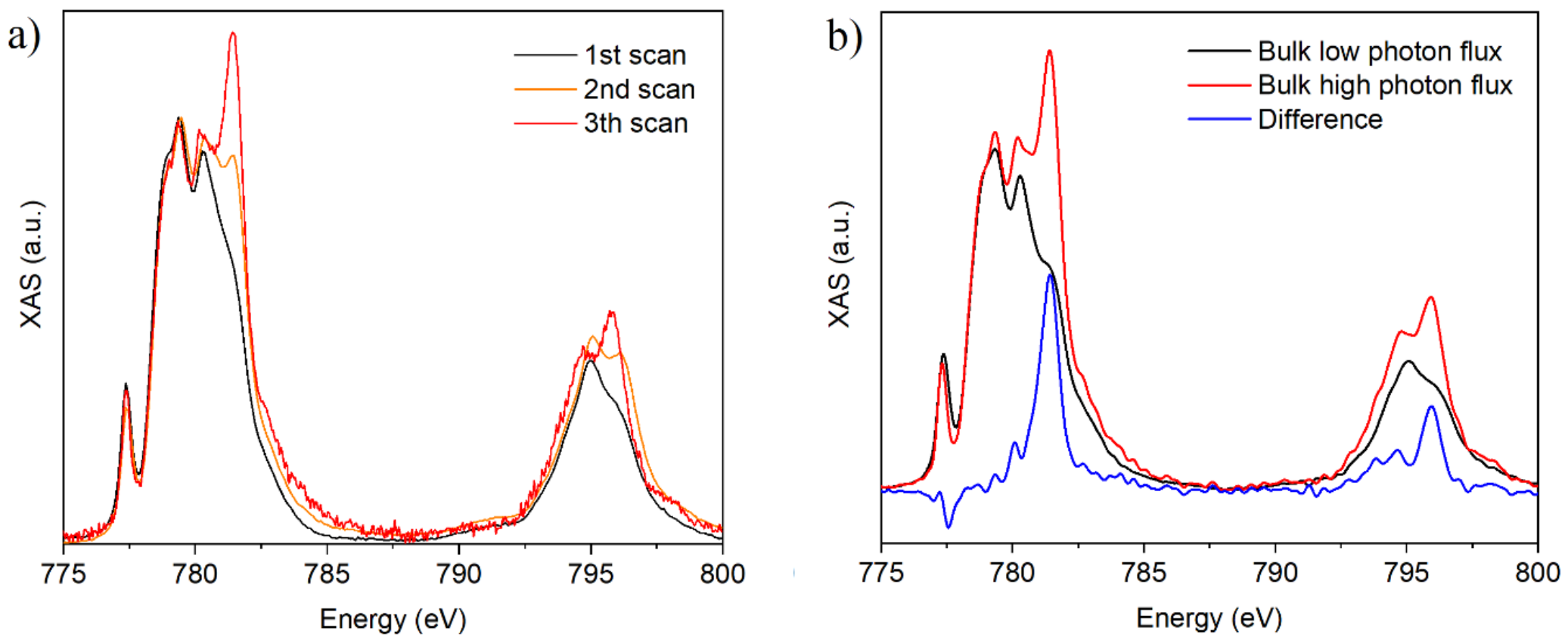

**Figure S.14. a)** Time evolution of Co$L_{2,3}$-edge XAS spectra of **1** at 300K with high photon flux and **b)** with different photon flux intensities and the calculated difference.

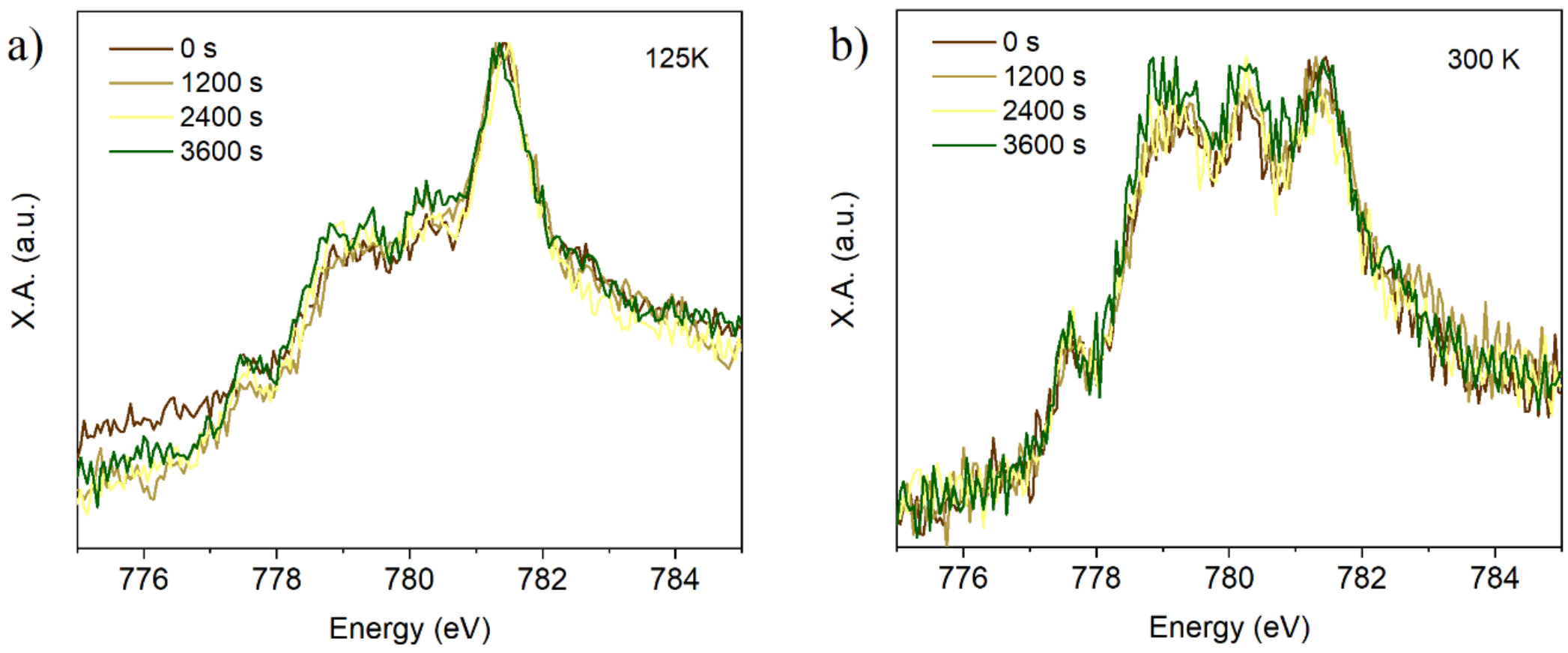

**Figure S.15.** Time evolution of Co$L_{2,3}$-edge XAS spectra of **1·SAM** at 100 K and 300 K.

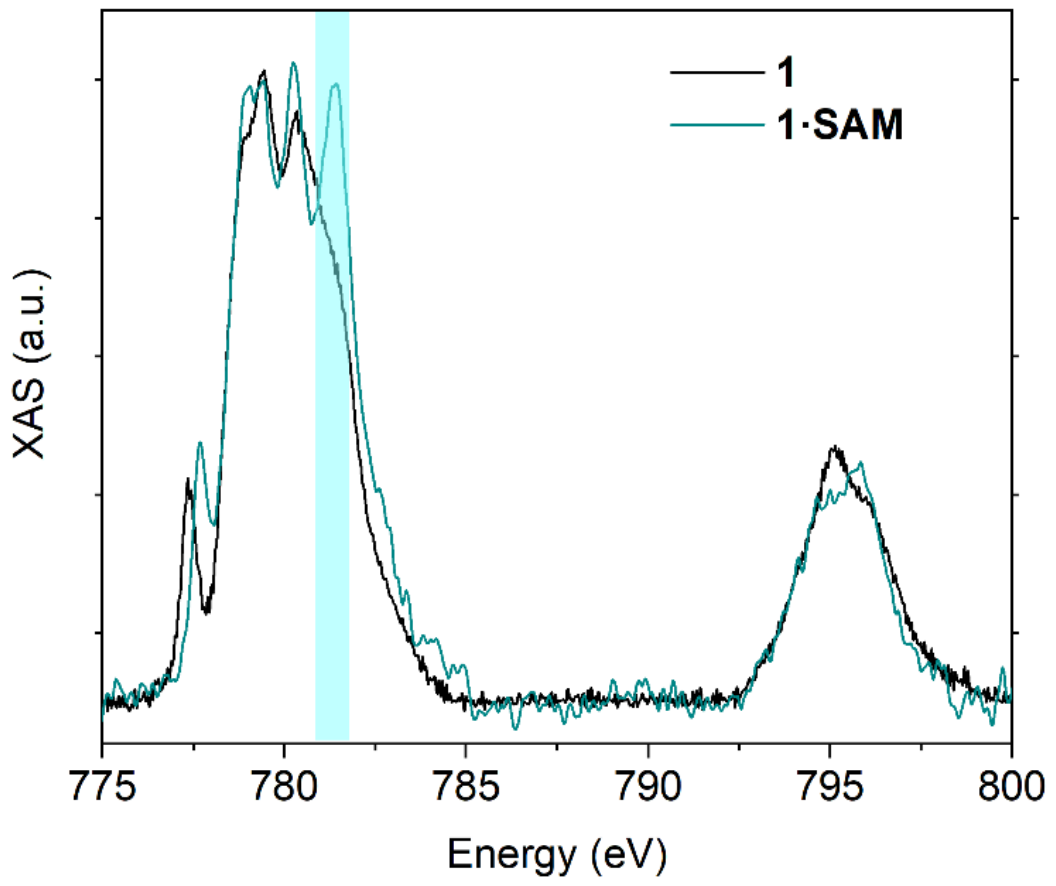

**Figure S.16.** Co$L_{2,3}$-edge XAS spectra of **1** (black line) and **1·SAM** (green line) at 300 K. Highlighted in cyan the peak at *ca.* 781.4 eV.

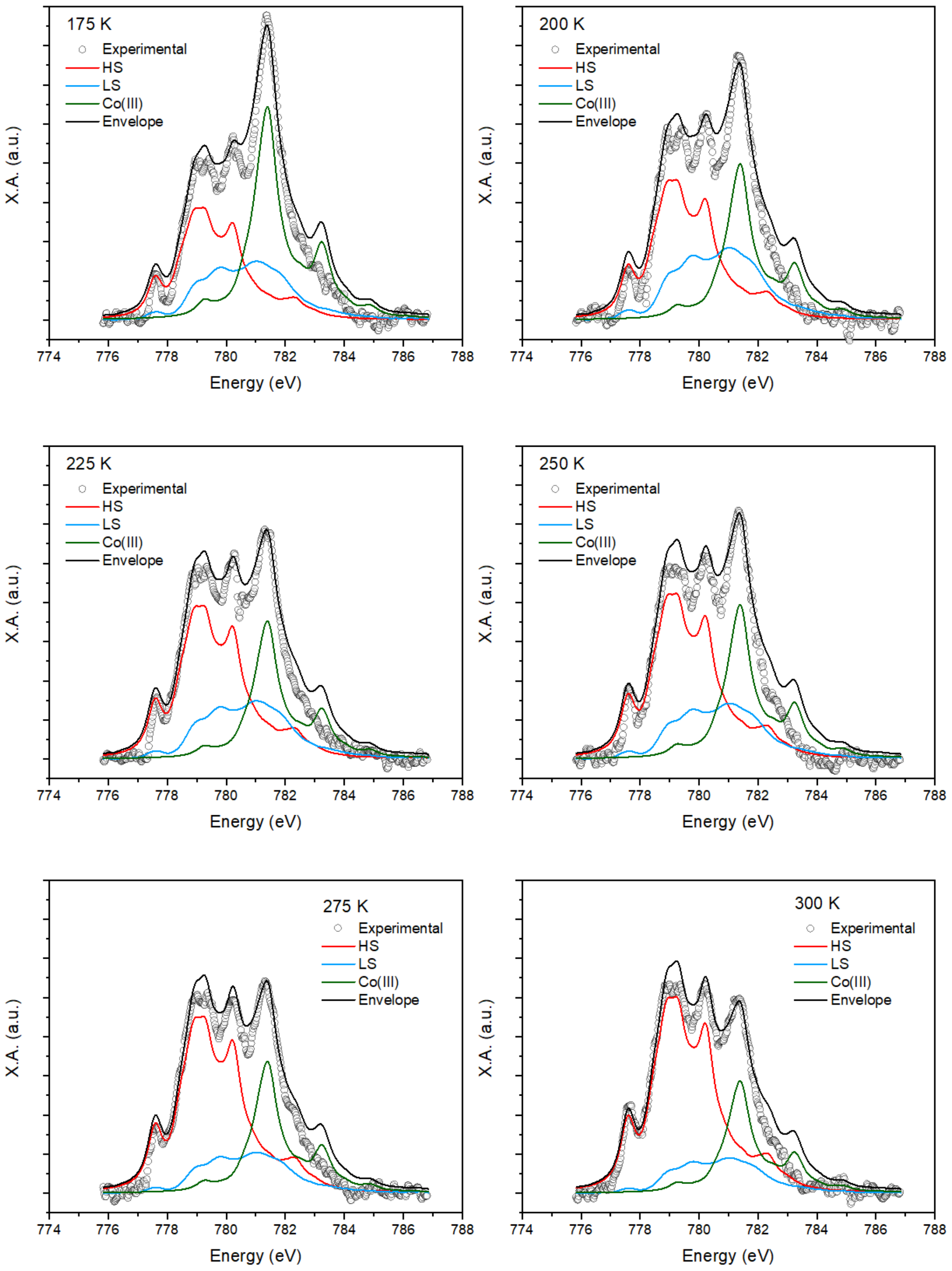

**Figure S.17.** Experimental XAS Co$L_3$-edge of **1·SAM** (empty circles) at each temperature of the 1st warming mode together with the theoretical HS Co(II) (red curves) and LS Co(III) (green curve) and experimental LS Co(II) (blue curves) references used for the fitting. Black line is the best fitting curve with the sum of the three components.

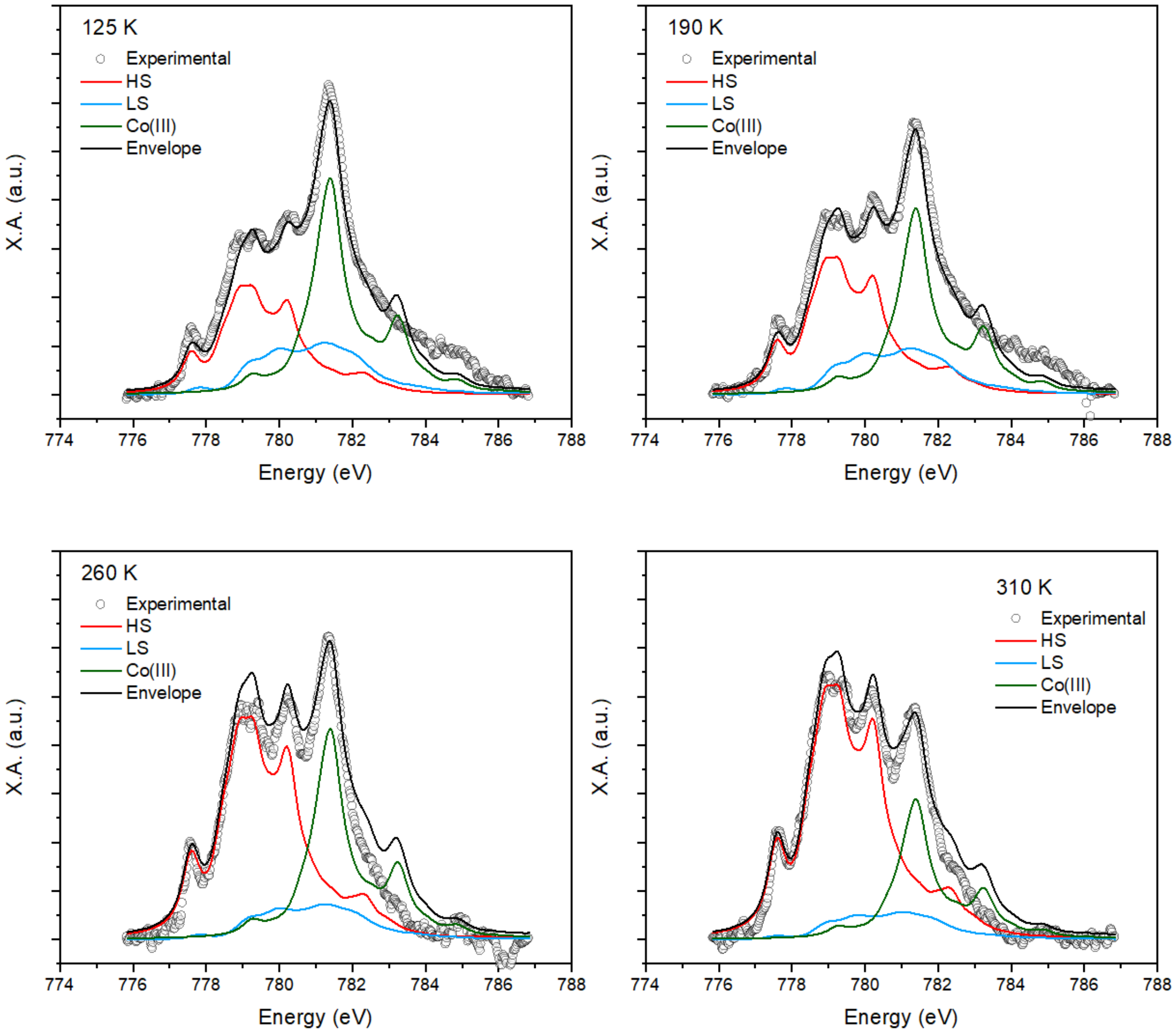

**Figure S.18.** Experimental XAS Co L3-edge of **1·SAM** (empty circles) at each temperature of the 2nd warming mode together with the theoretical HS Co(II) (red curves) and LS Co(III) (green curve) and experimental LS Co(II) (blue curves) references used for the fitting. Black line is the best fitting curve with the sum of the three components.

**Table S.7.** Experimental fraction of the different spin and valence states of Co molecules extracted from the fit.

| Temperature (K) | Co(II)-HS | Co(II)-LS | Co(III)-LS |
|---|---|---|---|
| 175 | 0.43 (0.02) | 0.20 (0.02) | 0.37 (0.02) |
| 200 | 0.50 (0.03) | 0.24 (0.02) | 0.26 (0.02) |
| 225 | 0.57 (0.02) | 0.20 (0.01) | 0.24 (0.01) |
| 250 | 0.58 (0.03) | 0.18 (0.02) | 0.25 (0.01) |
| 275 | 0.64 (0.02) | 0.14 (0.01) | 0.22 (0.01) |
| 300 | 0.70 (0.03) | 0.11 (0.01) | 0.18 (0.01) |
| 125 | 0.39 (0.02) | 0.25 (0.02) | 0.36 (0.02) |
| 190 | 0.48 (0.02) | 0.22 (0.02) | 0.30 (0.01) |
| 260 | 0.61 (0.02) | 0.13 (0.02) | 0.26 (0.01) |
| 310 | 0.74 (0.02) | 0.07 (0.01) | 0.19 (0.01) |
| 370[a] | 0.88 (0.02) | 0.06 (0.01) | 0.06 (0.01) |

**Table S.8.** Experimental fraction of Co(II)-HS/LS molecules extracted from the fit.

| Temperature (K) | Co(II)-HS | Co(II)-LS |
|---|---|---|
| 175 | 0.67 (0.04) | 0.33 (0.03) |
| 200 | 0.68 (0.04) | 0.32 (0.03) |
| 225 | 0.74 (0.03) | 0.26 (0.02) |
| 250 | 0.77 (0.04) | 0.23 (0.02) |
| 275 | 0.83 (0.03) | 0.17 (0.01) |
| 300 | 0.86 (0.03) | 0.14 (0.02) |
| 125 | 0.61 (0.03) | 0.39 (0.03) |
| 190 | 0.69 (0.03) | 0.31 (0.03) |
| 260 | 0.83 (0.03) | 0.17 (0.02) |
| 310 | 0.91 (0.03) | 0.09 (0.01) |
| 370[a] | 0.94 (0.02) | 0.6 (0.01) |

a. The temperature 370K was not consider due to the blocking of the spin state.

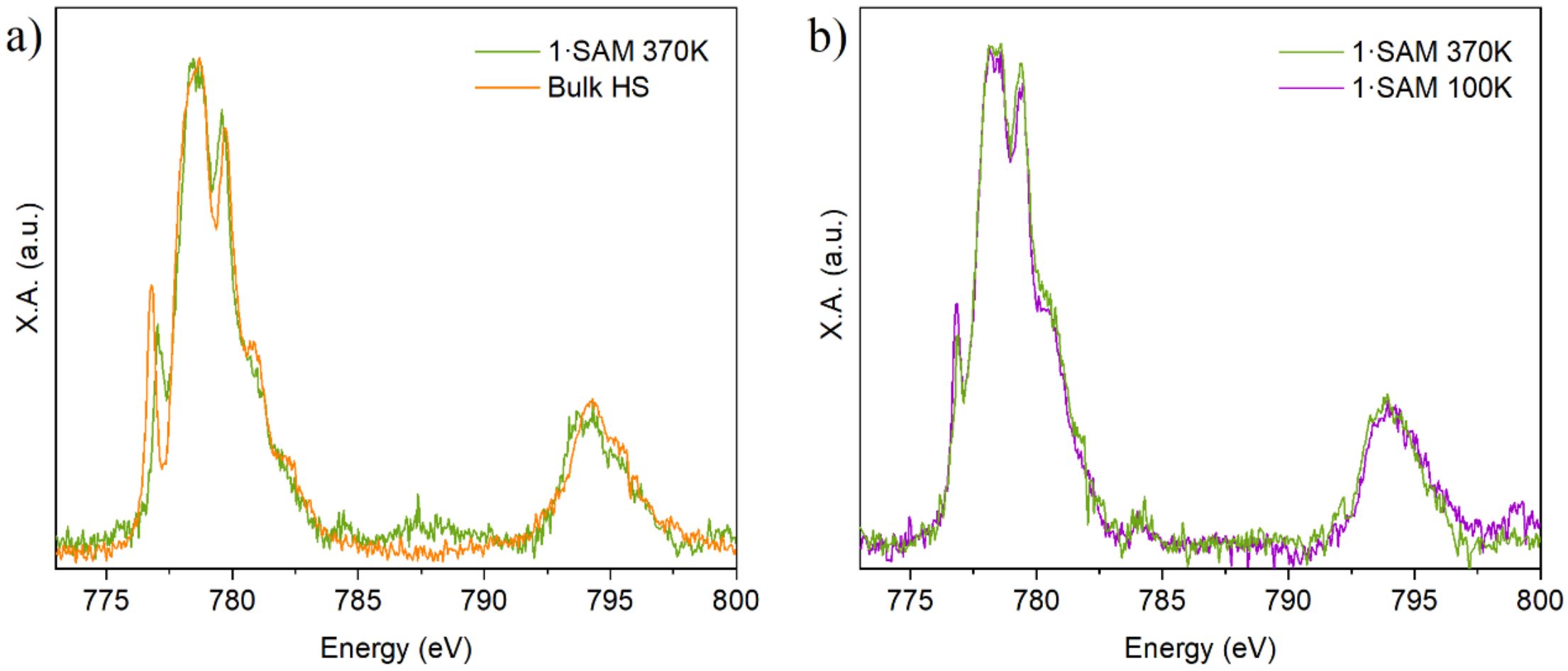

**Figure S.19.** Comparison of Co $L_{2,3}$-edge XAS spectra: **a)** at T=370K: **1·SAM** and bulk HS (raw spectrum subtracted from a 0.4 fraction of the spectrum measured at 100K), **b) 1·SAM** at 370K and its subsequent cooling down to 100K.